\documentclass[twocolumn]{aastex631}
\usepackage{amsmath}
\usepackage{graphicx}
\graphicspath{{./}{figures/}}

\usepackage{soul}

\newcommand{\Ha}{H$\alpha$}
\newcommand{\MgII}{Mg~\textsc{ii}}
\newcommand{\SiIV}{Si~\textsc{iv}}
\newcommand{\HRIEUV}{HRI$_{\mathrm{EUV}}$}
\newcommand{\kms}{\mathrm{km\,s^{-1}}}
\newcommand{\Mm}{\mathrm{Mm}}

\newif\ifdraftnobib
\draftnobibfalse
\ifdraftnobib
  \AtBeginDocument{%
    \renewcommand{\citep}[1]{\textit{[citation]}}%
    \renewcommand{\citet}[1]{\textit{[citation]}}%
  }
\fi

\newcommand{\figonefile}{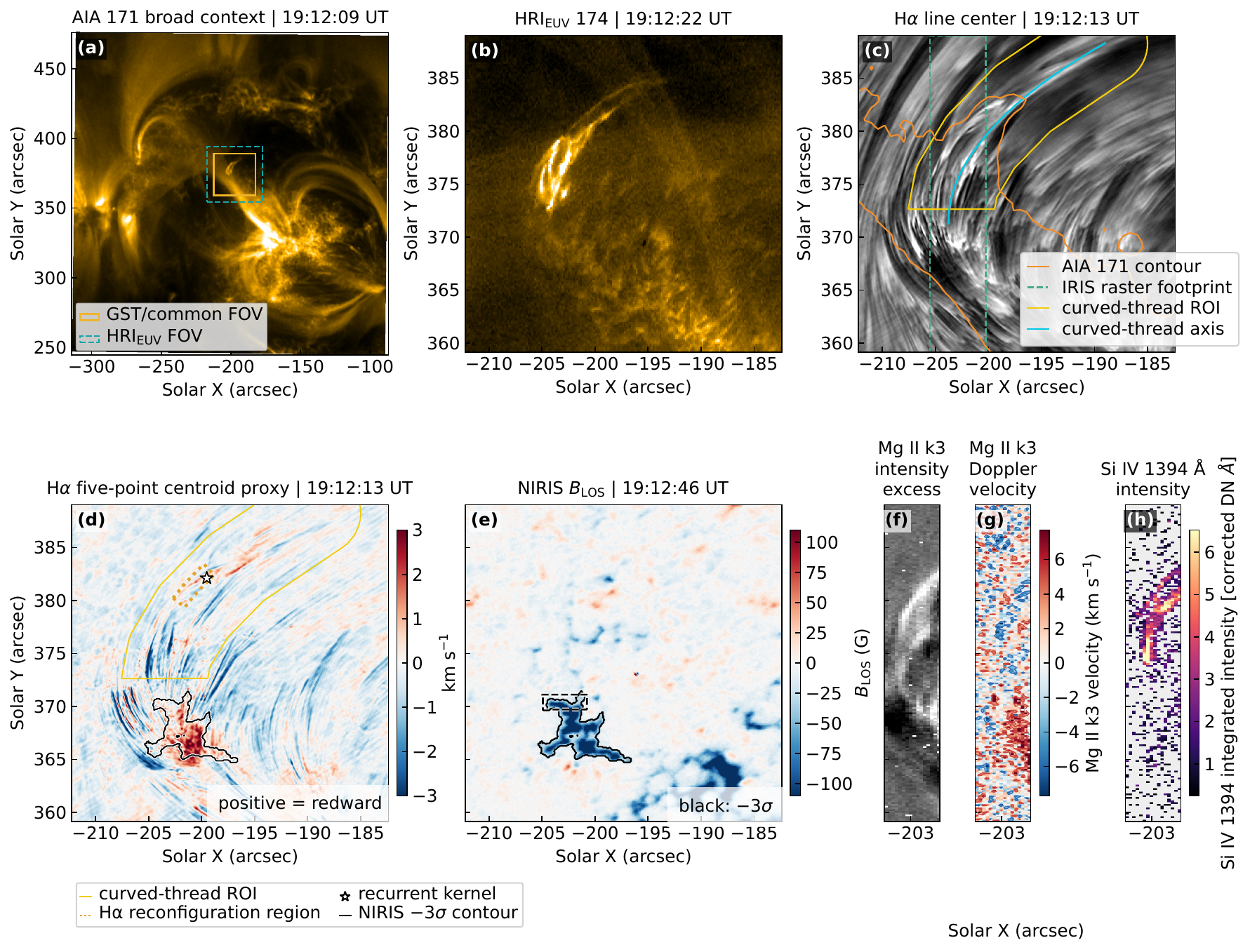}
\newcommand{\figtwofile}{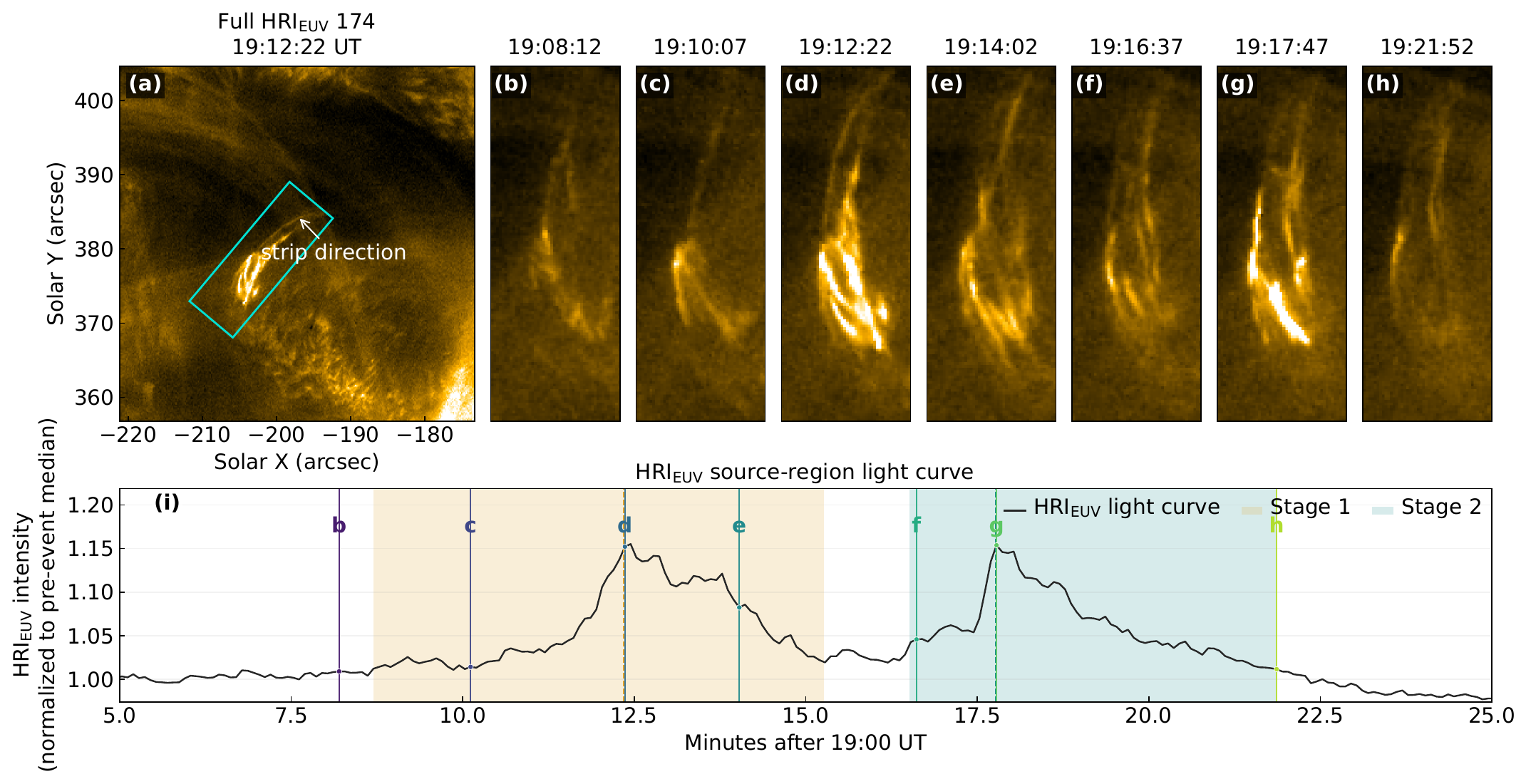}
\newcommand{\figthreefile}{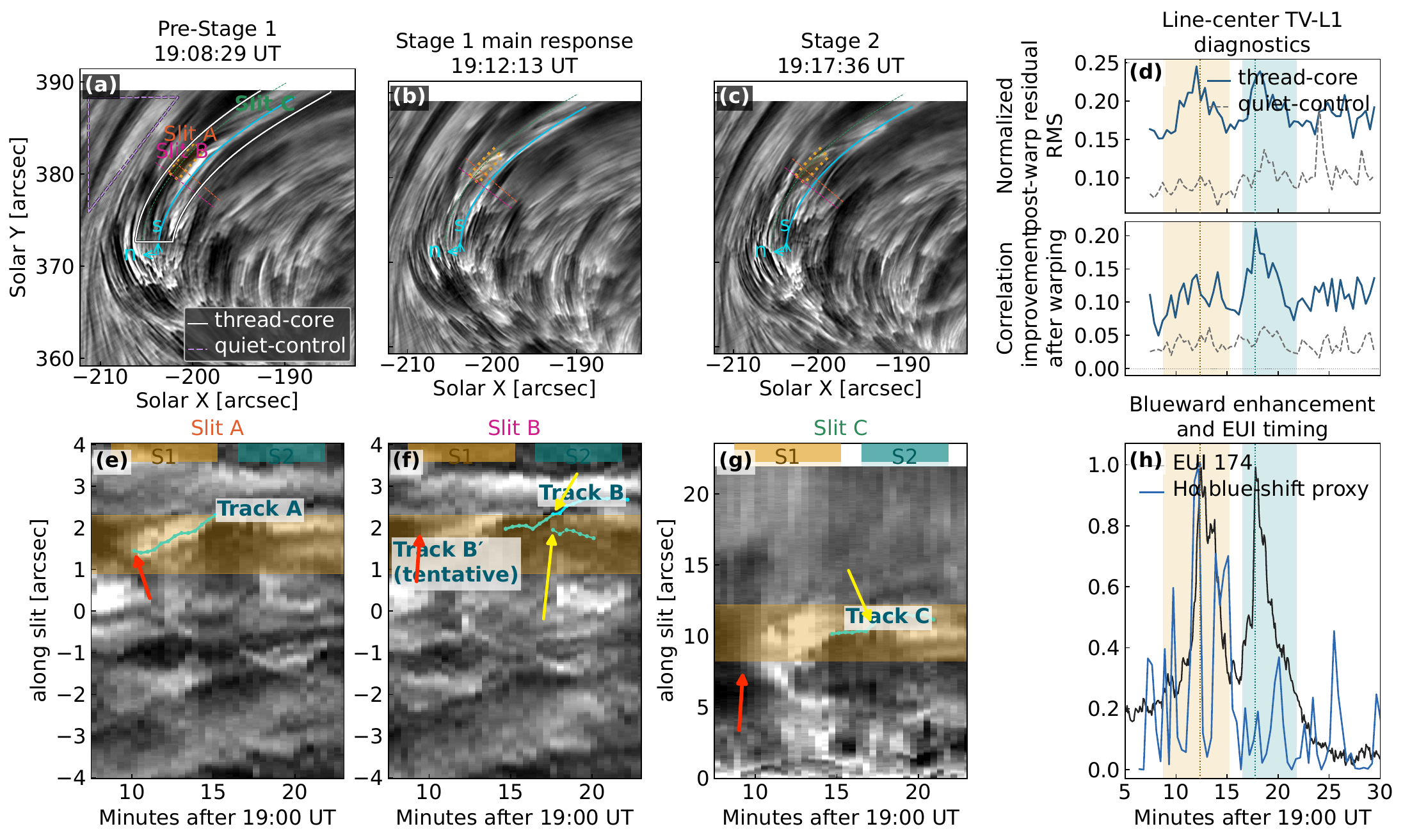}
\newcommand{\figfourfile}{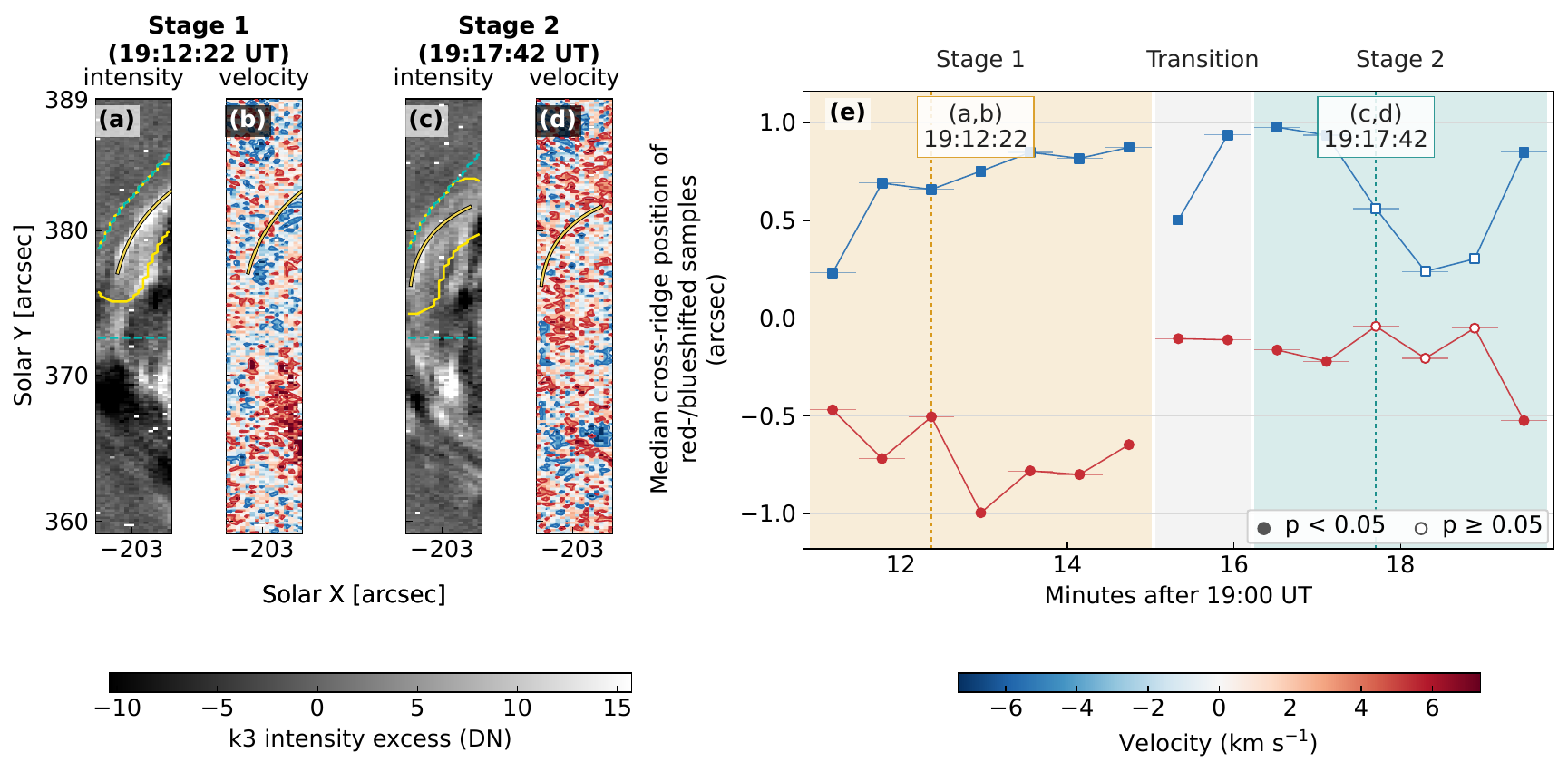}
\newcommand{\figfivefile}{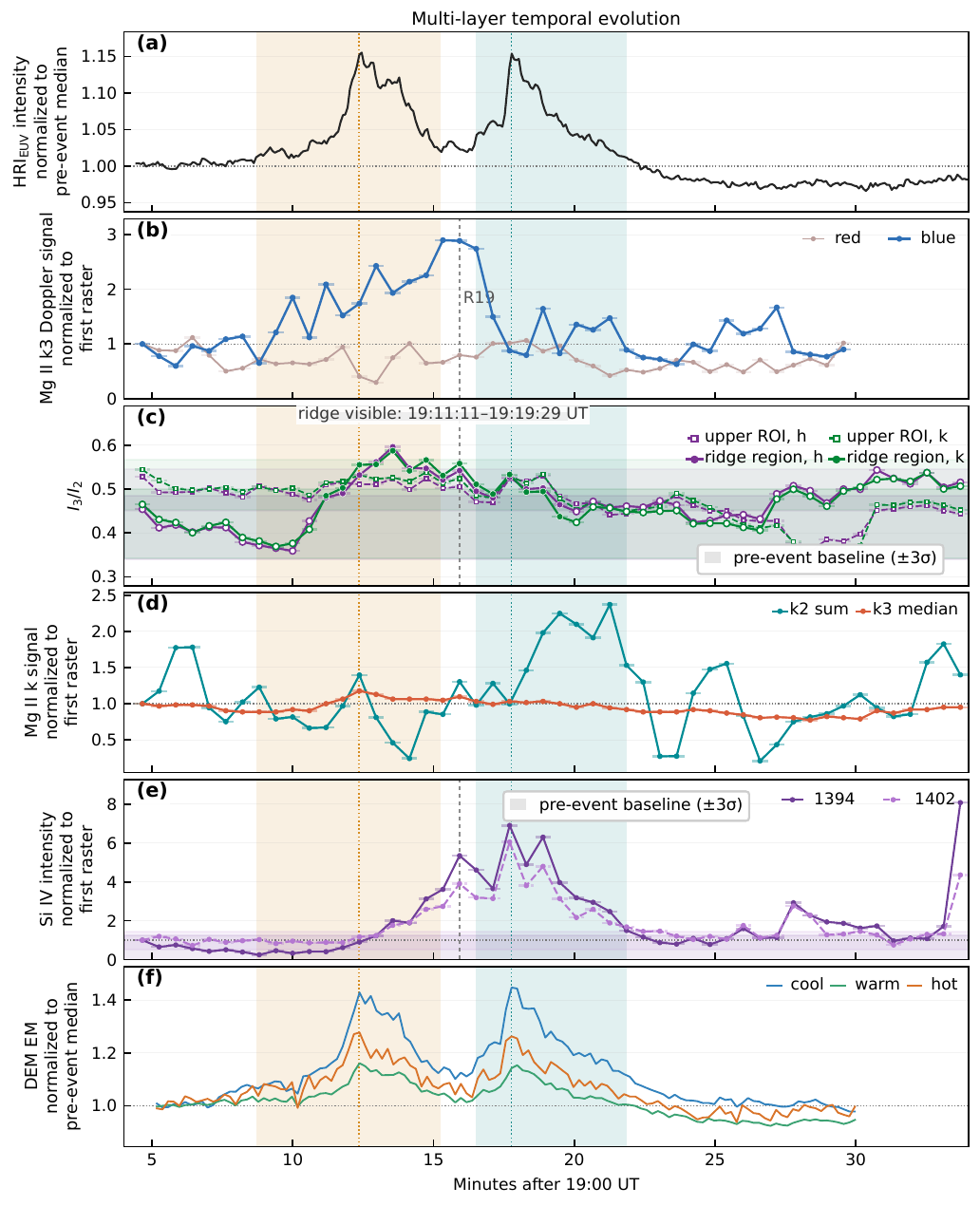}
\newcommand{\figsixfile}{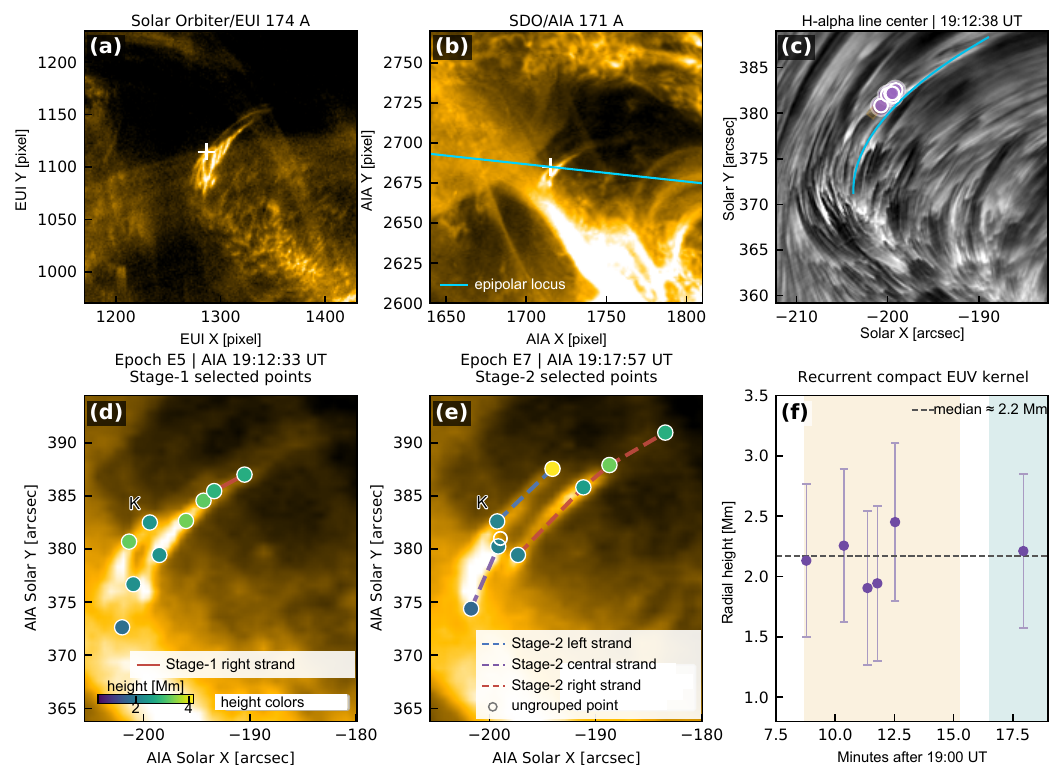}
\newcommand{\figsevenfile}{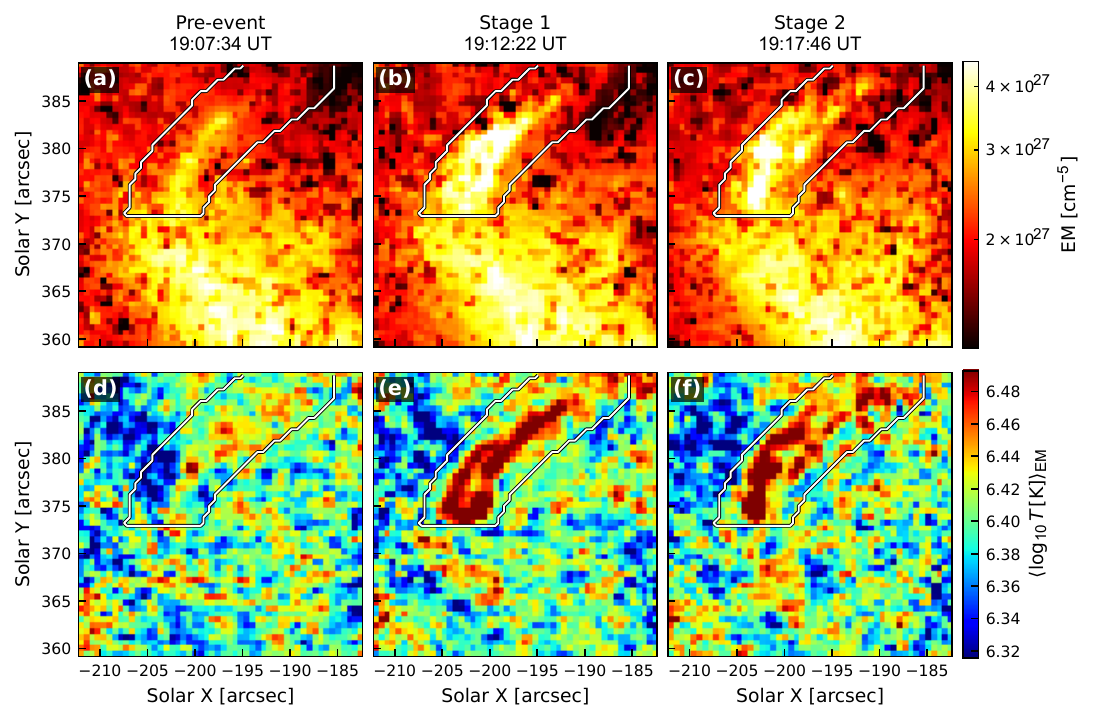}
\newcommand{\figsalignfile}{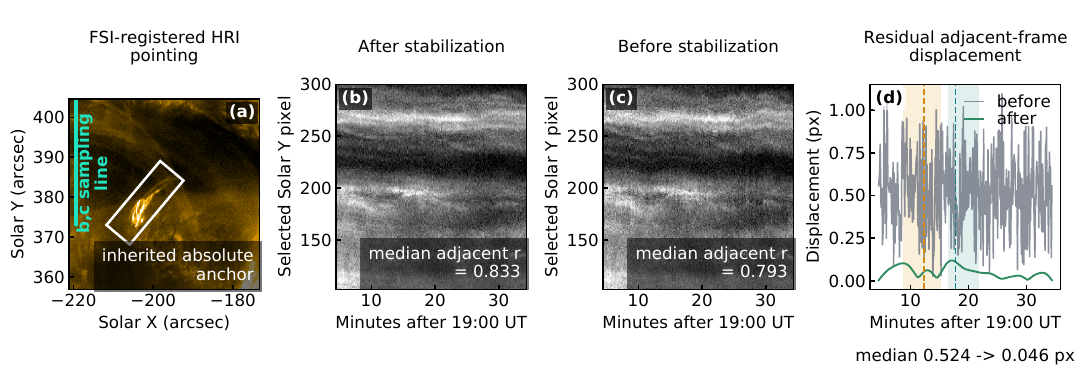}
\newcommand{\figsstereoafile}{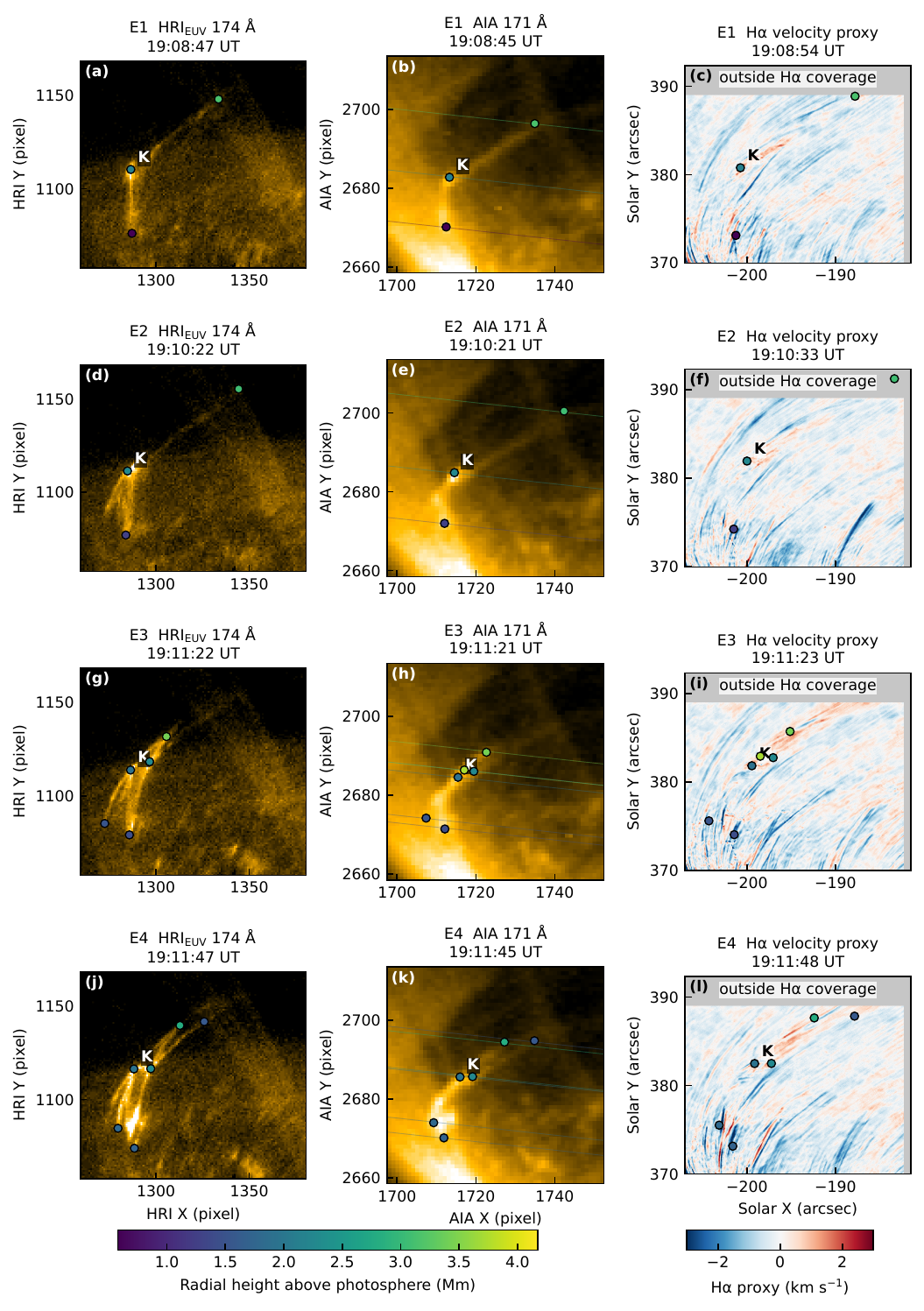}
\newcommand{\figsstereobfile}{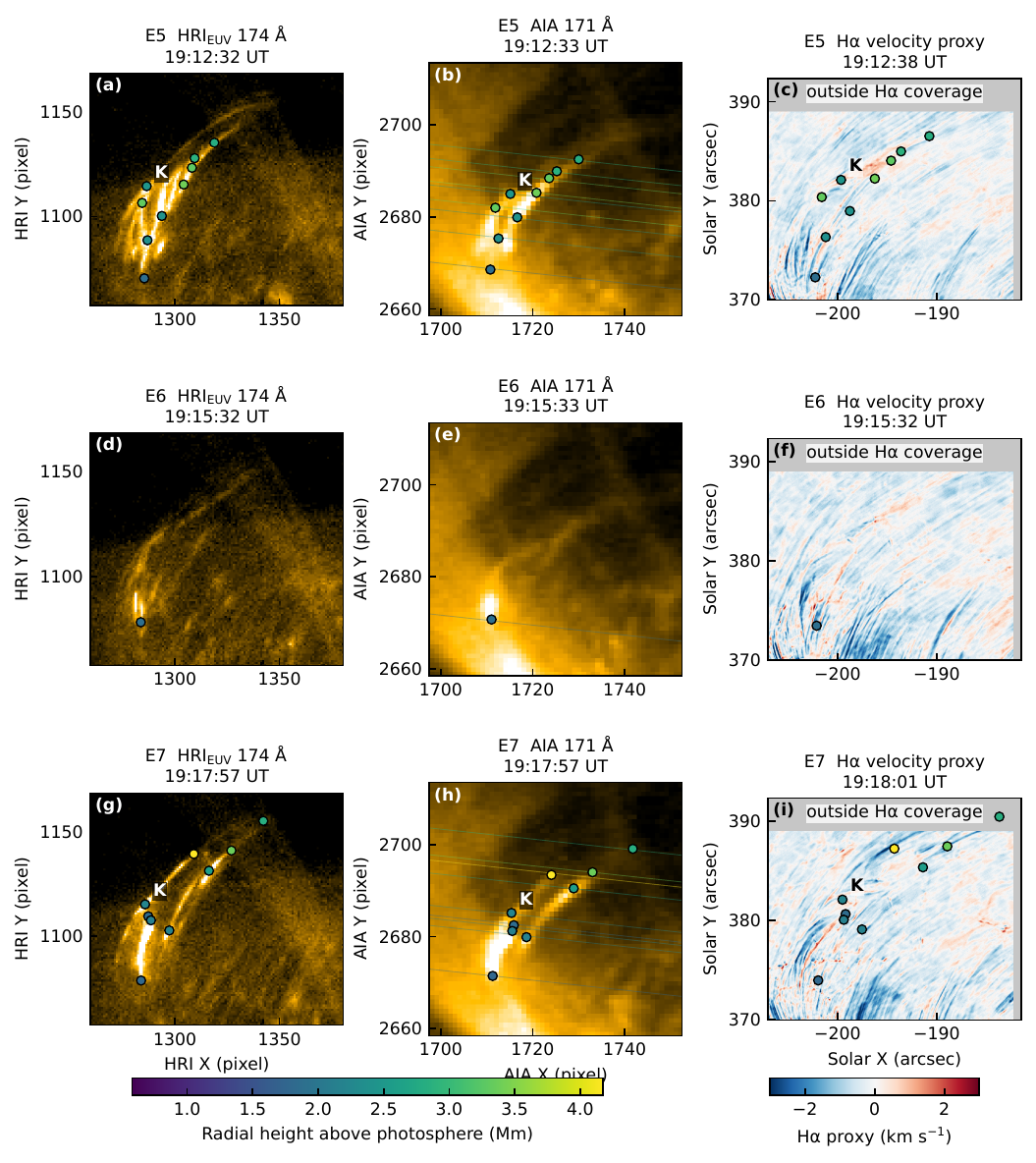}
\newcommand{\figsmultilayerfile}{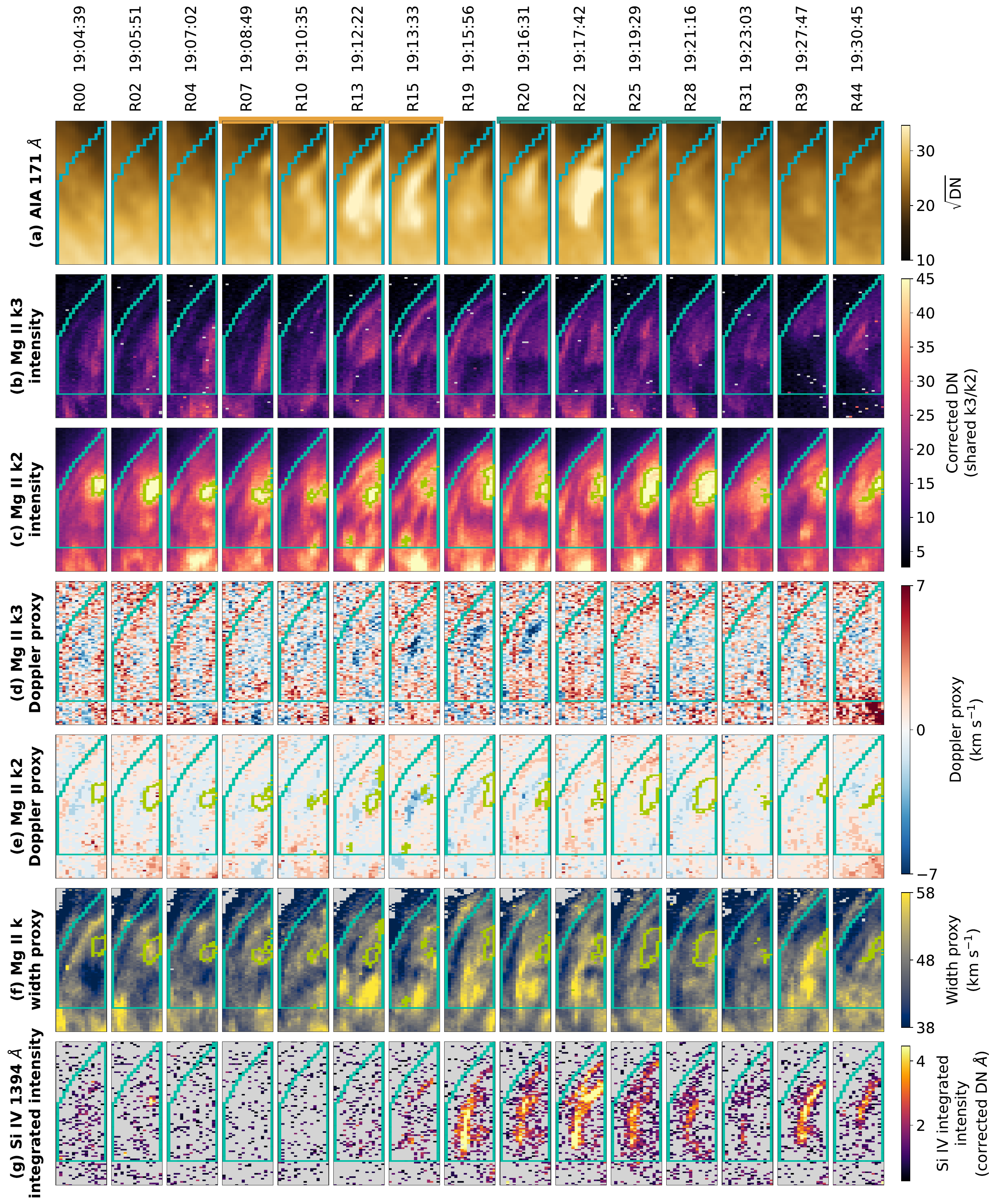}

\newcommand{\paperfigure}[4]{%
\begin{figure*}[t]
\centering
\includegraphics[width=#2\textwidth]{#1}%
\caption{#3}
\label{#4}
\end{figure*}}

\newcommand{\paperanimation}[5]{%
\begin{figure*}[t]
\centering
\includegraphics[width=#3\hsize]{#2}%
\caption{#4}
\label{#5}
\end{figure*}}

\submitjournal{ApJ}
\shorttitle{Tiny quiet-Sun EUV ejections}
\shortauthors{Huang et al.}

\begin{document}

\title{Tiny Quiet-Sun EUV Ejections Revealed by High-resolution Multiwavelength Observations}

\author[0000-0001-9049-0653]{Nengyi Huang}
\affiliation{Institute for Space Weather Sciences, New Jersey Institute of Technology, 323 Martin Luther King Blvd, Newark, NJ 07102-1982, USA}

\author[0000-0002-6628-6211]{Yuqian Wei}
\affiliation{Institute for Space Weather Sciences, New Jersey Institute of Technology, 323 Martin Luther King Blvd, Newark, NJ 07102-1982, USA}

\author[0000-0002-3669-1830]{Qin Li}
\affiliation{Institute for Space Weather Sciences, New Jersey Institute of Technology, 323 Martin Luther King Blvd, Newark, NJ 07102-1982, USA}

\author[0000-0002-5865-7924]{Jeongwoo Lee}
\affiliation{Institute for Space Weather Sciences, New Jersey Institute of Technology, 323 Martin Luther King Blvd, Newark, NJ 07102-1982, USA}

\author[0000-0002-5233-565X]{Haimin Wang}
\affiliation{Institute for Space Weather Sciences, New Jersey Institute of Technology, 323 Martin Luther King Blvd, Newark, NJ 07102-1982, USA}
\affiliation{Big Bear Solar Observatory, New Jersey Institute of Technology, 40386 North Shore Lane, Big Bear City, CA 92314-9672, USA}

\begin{abstract}
We investigate two successive tiny quiet-Sun EUV ejections on 2022 October 29 using coordinated Solar Orbiter/EUI, IRIS, SDO/AIA, and Goode Solar Telescope observations. We examine their low-lying geometry and the chromospheric and transition-region response before and between the releases. A curved H$\alpha$ thread is present beforehand, and \MgII\ and \SiIV\ spectra reveal a compact multithermal precursor 2.6--2.7~min before the first EUV brightening. EUI--AIA stereoscopy places the recurrent compact EUV kernel near 2.2~Mm and the sampled emission at approximately 0.6--4.2~Mm along inclined, locally arched structures. The first ejection shows multithreaded, helical-looking EUV strands; the second shows straighter strands and a decrease in the cross-ridge separation of redshifted and blueshifted \MgII\ k3 samples from $1.47''$ to $0.89''$. Nevertheless, the two peaks have comparable emission-measure increments and estimated event-associated thermal contents of a few $10^{26}$~erg under the same assumptions. Before the second ejection, the regional \SiIV\ and k3 blueshift signals remain enhanced, while the \MgII\ central reversal along the activated ridge remains shallower than before the event. Component reconnection between the curved structure and the surrounding field offers a possible explanation. The observations show that jet-like EUV emission can occur in low-lying structures. The second ejection begins before the chromospheric and transition-region spectra have returned to their pre-event states. Its spectral signatures may therefore depend on the atmospheric changes associated with the first ejection as well as on the additional energy input.
\end{abstract}

\section{Introduction}
\label{sec:introduction}

Small-scale jets and jet-like brightenings are ubiquitous signatures of impulsive magnetic-energy release in the solar atmosphere.
They span classical coronal-hole and active-region jets, jetlets, network jets, campfires, and picoflare-scale ejections \citep{raouafi_solar_2016,sterling_small-scale_2015,panesar_iris_2018,berghmans_extreme-uv_2021,chitta_picoflare_2023,huang_statistical_2023,huang_high-resolution_2025}.
Their occurrence rates make them plausible contributors to coronal heating and, in open-field environments, to the mass, momentum, waves, and magnetic stress carried by the solar wind.
Solar Orbiter (SolO; \citealt{muller_solar_2020}) observations have connected widespread picoflare jets in coronal holes with open magnetic flux and in-situ solar-wind streams \citep{chitta_picoflare_2023,chitta_coronal_2025}.
Understanding the broader population also requires measurements of the pre-existing lower-atmosphere structures, their response to individual releases, and the geometry of the resulting emission.

A collimated extreme-ultraviolet (EUV) feature can trace bulk flow, a propagating heating front, sequential strand illumination, or a mixture of these processes.
Some quiet-Sun EUV brightenings are located only 1--5~Mm above the photosphere and are consistent with internally heated small loops \citep{zhukov_stereoscopy_2021,chen_transient_2021}.
Jets can also be redirected or confined by overlying loops, with cool and hot components following related but non-identical paths \citep{yang_observational_2023}.
For such low-lying events, the geometry of the emitting strands must be considered together with the evolution of the cool structure and its chromospheric and transition-region (TR) counterparts.
The resulting combination constrains how a jet-like coronal response develops within, and interacts with, an already structured lower atmosphere.

Twist provides a natural energy reservoir for many jets.
In interchange-reconnection and blowout-jet models, reconnection transfers twist from a closed system into longer field lines and launches an untwisting or torsional-Alfv\'enic disturbance \citep{Pariat2009,Pariat2015,Pariat2016,sterling_small-scale_2015}.
High-resolution observations have shown reconnection between erupting filament threads and neighboring chromospheric fibrils, followed by connectivity change and twist release \citep{xue_observing_2016}.
Related events involve repeated interaction of a chromospheric flux rope with ambient field, external reconnection between a filament and adjacent loops, and opposite-sided Mg~\textsc{ii} Doppler signals accompanying resolved untwisting \citep{kumar_formation_2015,li_magnetic_2016,panesar_buildup_2025}.
Here, component reconnection denotes reconnection between obliquely oriented fields that retain a finite guide-field component.

The morphology of a single magnetic system can differ strongly across atmospheric diagnostics.
\Ha\ line-center fibrils are optically thick structures whose visibility follows the evolving optical-depth-unity surface, mass loading, velocity, and foreground absorption; a visible fibril need not trace one fixed field line for its entire lifetime \citep{leenaarts_fibrils_2015,pereira_appearance_2016}.
EUV emission is optically thin and temperature selective, so it highlights only heated portions of the same field-aligned system and integrates multiple structures along the line of sight (LOS).
Coordinated observations show that chromospheric fibrils or spicules can acquire TR and coronal counterparts with offsets, different widths, and different apparent shapes as only selected threads or tips reach hotter temperature regimes \citep{rouppe_van_der_voort_heating_2015,mandal_evolution_2023}.
Thus, spatial correspondence and coherent timing can support an association with a common evolving system even when \Ha\ and EUV images do not reproduce the same projected outline.

The Interface Region Imaging Spectrograph (IRIS) adds height-sensitive spectroscopic information to this comparison.
The \MgII\ k2 peak intensities provide a temperature-sensitive diagnostic of the middle chromosphere, particularly for strong emission peaks, while k3 generally samples the upper chromosphere nearer the TR and its Doppler shift follows the velocity near the k3 optical-depth-unity surface \citep{leenaarts_formation_2013}.
Compact Mg~\textsc{ii} and \SiIV\ emission together with broad or multicomponent profiles and opposite-signed velocities can mark lower-atmosphere reconnection before an overlying EUV response becomes prominent \citep{li_spectroscopic_2018}.
Other small jets evolve from a mild cool \Ha\ phase into a hotter EUV phase as the dissipation region develops \citep{tang_photospheric_2025}.
These diagnostics make it possible to reconstruct the buildup, staged release, and atmospheric aftermath of a tiny event.

Recent perihelion observations with the Extreme Ultraviolet Imager (EUI) on SolO, particularly its High Resolution Imager in the EUV channel (\HRIEUV), resolve fine structure in very small EUV ejections \citep{chitta_picoflare_2023}. \citet{petrova_observations_2024} reported propagating rotational motions in a fan--spine structure observed with SolO/EUI, consistent with torsional Alfv\'en waves or flux-rope untwisting.

\citet{lee_fine_2025} studied the present event using EUI and the Goode Solar Telescope (GST), including \Ha\ observations with the Visible Imaging Spectrometer (VIS) and magnetograms from the Near InfraRed Imaging Spectropolarimeter (NIRIS). They resolved EUV strands about 0.7~Mm wide and interpreted the first ejection in terms of interchange reconnection. Using a reconnection model \citep{longcope_alfven_2025}, they estimated a magnetic-energy release of order $10^{26}$~erg and found broad agreement with the estimated jet energy, dominated by its thermal component. For their differential emission measure (DEM) analysis, they used the deep-learning method of \citet{youn_can_2025} to generate images equivalent to channels of the Atmospheric Imaging Assembly (AIA) on the Solar Dynamics Observatory (SDO). These AIA-equivalent images, together with the observed EUI emission, provided the thermal diagnostic.

In this paper, we study the same event with wider temporal coverage, IRIS ultraviolet (UV) spectroscopy, and stereoscopy combining \HRIEUV\ with SDO/AIA. We address three questions. First, what chromospheric structure and activity precede the first release? Second, how do the chromospheric and TR diagnostics evolve during and between the releases, and has their response subsided before the second release begins? Third, at what heights is the jet-like EUV emission located, and how do the thermal responses of the two releases compare? Together, these measurements constrain the geometry of the emitting structures and the atmospheric conditions in which successive ejections occur.

\section{Observations and Data Analysis}
\label{sec:data}

The coordinated data set combines \HRIEUV\ 174~\AA\ imaging, GST/VIS \Ha\ imaging spectroscopy, GST/NIRIS magnetic-field measurements, IRIS \MgII\ and \SiIV\ spectra, and SDO/AIA imaging and DEM diagnostics.
Table~\ref{tab:sampling} summarizes the temporal sampling and time representation adopted for each instrument.

Figure~\ref{fig:context} summarizes the multi-instrument context and field-of-view (FOV) coverage. All times are given in Universal Time (UT). For \HRIEUV, we use the Earth-arrival time provided with each EUI image, so that the EUI and Earth-based observations refer to the same time of light arrival at Earth; for the other instruments, we use the exposure times.

The region of interest (ROI) follows the curved \Ha\ structure and is outlined in yellow in Figure~\ref{fig:context}(c,d). We use its upper portion, hereafter the \emph{upper ROI}, to include the structures associated with the EUV activity while excluding the strong signal near the photospheric footpoint. Regional IRIS measurements use the part of the upper ROI covered by the raster, outlined in teal in Figure~\ref{fig:supp_multilayer} and by dashed teal boundaries in Figure~\ref{fig:mg}(a,c). We use \emph{EUV source region} for the area occupied by the event-related EUV brightening in Figure~\ref{fig:eui}; the recurrent compact EUV kernel is a localized bright feature within this region.

\begin{deluxetable*}{lll}
\tablewidth{0pt}
\tablecaption{Temporal sampling and time representation used in this study.
\label{tab:sampling}}
\tablehead{\colhead{Instrument/channel} &
\colhead{Acquisition} &
\colhead{\shortstack{Time\\representation}}}
\startdata
\HRIEUV\ 174~\AA & 5.0~s image cadence & Individual Earth-arrival times \\
GST/VIS \Ha & \shortstack[l]{Five sequential wavelength positions;\\7.2~s scan; 24.9~s line-center repeat} & \shortstack[l]{Individual exposure times;\\scan interval retained} \\
IRIS \MgII/\SiIV & \shortstack[l]{16-step raster; 33.3~s duration;\\35.6~s midpoint cadence} & \shortstack[l]{Per-step exposure times and\\raster start--end intervals} \\
GST/NIRIS & 51.0~s cadence & Individual exposure times \\
SDO/AIA & 12.0~s image cadence & Individual exposure times \\
\enddata
\tablecomments{Each IRIS raster is a sequentially sampled space--time mosaic rather than an instantaneous two-dimensional image. The rasters are numbered R00--R49 in temporal order.}
\end{deluxetable*}

\paperfigure{\figonefile}{0.99}
{Spatial coverage of the instruments.
(a) AIA 171~\AA\ broad context. (b) \HRIEUV\ 174~\AA\ local context.
(c) Co-aligned \Ha\ line-center image; the yellow outline marks the curved-thread ROI and the cyan curve its adopted axis. The orange contour provides AIA 171~\AA\ context, and the outlined footprint marks the IRIS raster coverage.
(d) \Ha\ five-point centroid-velocity proxy in the same registered FOV, with the ROI, amber dotted reconfiguration region, recurrent EUV kernel (star), and NIRIS negative-field contour. Positive proxy values denote redward shifts.
(e) NIRIS LOS magnetic field, $B_{\rm LOS}$, at 19:12:46~UT. The dashed dark boundary encloses the fixed compact subregion used for the magnetic-flux statistics; the solid black contour marks the connected negative-field structure at $-3\sigma$ and is not the statistical-region boundary.
(f), (g) \MgII\ k3 intensity excess and Doppler velocity for the Stage~1 raster R13, labeled S1, centered at 19:12:22~UT and acquired over 19:12:05--19:12:39~UT.
(h) \SiIV\ 1394~\AA\ intensity for the Stage~2 raster R22, labeled S2, centered at 19:17:42~UT and acquired over 19:17:26--19:17:59~UT. These are representative scans from different stages, not simultaneous snapshots.}
{fig:context}

\subsection{Event timing and common coordinate system}
\label{sec:timing}

Elapsed times in the figures are given in minutes after 2022 October 29 19:00:00~UT.
The two intervals were selected from the native-cadence source-region \HRIEUV\ light curve and verified in the resolved image sequence.
Stage~1 begins at 19:08:42~UT, peaks at 19:12:21~UT, and ends at 19:15:16~UT.
Stage~2 begins at 19:16:31~UT, peaks at 19:17:46~UT, and ends at 19:21:51~UT (Table~\ref{tab:stages}).
The interval boundaries describe the observed rise and decay phases rather than formal change points. \par Throughout the time-series figures, pale amber shading denotes the Stage~1 interval and pale teal/cyan shading denotes Stage~2; Figure~\ref{fig:mg} additionally uses gray shading for the transition samples between them.

\begin{deluxetable}{lccc}
\tablewidth{0pt}
\tablecaption{Event intervals (UT).
\label{tab:stages}}
\tablehead{\colhead{Stage} & \colhead{Onset} & \colhead{\HRIEUV\ peak} & \colhead{End}}
\startdata
Stage 1 & 19:08:42 & 19:12:21 & 19:15:16 \\
Stage 2 & 19:16:31 & 19:17:46 & 19:21:51 \\
\enddata
\end{deluxetable}

The GST/VIS, IRIS, AIA, and NIRIS observations were registered in a common Earth-view coordinate system, while the \HRIEUV\ pointing was refined through registration to the EUI Full Sun Imager (FSI).
The final relative alignment accuracy is approximately one arcsecond or better.
Comparisons between subarcsecond features are therefore phrased in terms of the same EUV source region unless a tighter same-instrument correspondence is available.

\subsection{EUI and AIA imaging}
\label{sec:eui_data}

The primary coronal diagnostic is the SolO/EUI \HRIEUV\ 174~\AA\ sequence \citep{rochus_solar_2020} from EUI Data Release~6.0 (\dataset[https://doi.org/10.24414/z818-4163]{doi:10.24414/z818-4163}).
The analyzed interval spans 19:04:27--19:34:22~UT with a median cadence of 5.0~s.
Residual spacecraft jitter was removed and the absolute pointing was refined through \HRIEUV--FSI registration.
All event morphology, photometry, and stereoscopic tie points use the corrected sequence. The source-region light curve is obtained by summing the intensity within the teal strip outlined in Figure~\ref{fig:eui}(a).
AIA 171~\AA\ provides Earth-view coronal context at a 12~s cadence \citep{lemen_atmospheric_2012,Boerner2012}, and six AIA EUV channels provide the DEM diagnostics described below.

\subsection{GST/VIS \Ha\ imaging spectroscopy}
\label{sec:halpha_data}

GST/VIS sampled \Ha\ at line center and at \(\pm0.4\) and \(\pm0.8\)~\AA.
The line-center sequence has a median cadence of 24.9~s, and a complete five-position scan spans a median of 7.2~s.
All wavelength positions retain their individual exposure times.

The \Ha\ temporal measurements use the upper ROI defined above and shown within the yellow outline in Figure~\ref{fig:context}(c,d).
The adopted \Ha\ axis defines a local coordinate system: \(s\) is distance along the axis, \(L_{\rm axis}\) is its total length, and \(n\) is signed perpendicular distance, positive to the left of increasing \(s\). In the active part of the structure shown in Figure~\ref{fig:halpha}(a), positive \(n\) points toward the upper left.
We derive a five-point absorption-centroid velocity proxy relative to a fixed reference profile; positive values denote redward shifts.
The H$\alpha$ blueward-shift-strength proxy integrates reliable five-point centroid/pseudo-Doppler samples with \(v_{\rm c}<-3~\kms\) in the selected region.
Because \Ha\ is optically thick and the wavelength positions are sequential, this quantity is used as a centroid-shift proxy rather than a calibrated bulk-flow velocity.

Plane-of-sky evolution is measured along three fixed sampling paths, hereafter slits (Figure~\ref{fig:halpha}(a)): Slit A and Slit B are transverse cuts at $s/L_{\rm axis}=0.40$ and 0.35, respectively, each sampled over $n=-4''$ to $+4''$, and Slit C is a longitudinal path offset by $n=+1.8''$ from the \Ha\ axis. Their time--distance maps are shown in Figure~\ref{fig:halpha}(e)--(g).
A feature track is an intensity feature followed for as long as it remains identifiable in a time--distance map; its fitted slope gives an apparent pattern speed. Tracks A and B are transverse features in Slits A and B, Track B$'$ is a tentative counter-directed feature in Slit B, and Track C is a longitudinal feature in Slit C.
The \Ha\ reconfiguration region, outlined by the amber dotted boundary in Figures~\ref{fig:context}(d) and \ref{fig:halpha}(a) and shaded in Figure~\ref{fig:halpha}(b,c), marks the part of the line-center structure where pronounced changes are seen during the two releases. It is a spatial guide to the observed activity rather than an analysis mask.
To separate apparent displacement from other image changes, we also apply total-variation L1 (TV-L1) optical flow to consecutive line-center images. The earlier image is remapped according to the inferred displacements and compared with the later image. The normalized post-remapping root-mean-square (RMS) residual measures the image change not explained by apparent displacement, and the correlation gain measures the improvement in frame-to-frame similarity after remapping. These quantities are evaluated in the \Ha\ thread-core region, the part of the upper ROI within $-1.5''\le n\le+2.5''$ of the \Ha\ axis (white outline in Figure~\ref{fig:halpha}(a)), and in a quiet-control fibril region (purple dashed outline). They are image-comparison diagnostics, not plasma velocities; further details are given in Appendix~\ref{app:of}.

\subsection{IRIS \MgII\ and \SiIV\ Spectroscopy}
\label{sec:iris_data}

IRIS obtained 50 rasters, each with 16 slit steps, containing chromospheric \MgII\ h and k spectra and TR \SiIV\ 1394 and 1402~\AA\ spectra \citep{depontieu_interface_2014}.
Adjacent slit positions are separated by \(0.349\arcsec\) and sampled at a median interval of 2.17~s.
A complete raster spans \(5.24\arcsec\times29.87\arcsec\), has a median duration of 33.3~s, and repeats at a median midpoint cadence of 35.6~s.
Each complete raster is represented by the median of its 16 slit-step exposure times, referred to here as its midpoint and rounded to the nearest second in the text. Full start--end intervals are retained for timing-sensitive comparisons. Pixel-level precursor and boundary comparisons use the actual slit-step exposure times.

For self-reversed \MgII\ k profiles, k2V and k2R denote the blue and red emission peaks, and k3 denotes the central intensity minimum between them. We refer to this central absorption-like depression as the \emph{central reversal} and use k2 for the mean of the two peak intensities; h2 and h3 are defined analogously.
We use the IRIS\(^2\) database of representative profiles and model atmospheres \citep{sainz_dalda_recovering_2019} as an auxiliary tool. For complex profiles, the best-matching IRIS\(^2\) representative profile is used only to locate the expected h2, h3, k2, and k3 sampling positions; all intensities are measured from the observed spectrum.
The raster-level ratios defined below are medians over profiles for which both the central-reversal and emission-peak intensities can be measured.
Profiles that are too weak or too complex to provide these quantities are omitted from the median; their omission is not interpreted as a physical disappearance of the central reversal.
The profile-based k3 or h3 value is the observed three-sample median around the representative minimum.
The profile-based k2 or h2 value is the mean of the high-intensity blue and red regions in the adopted peak-search intervals after excluding the central-reversal samples, and is therefore a full-profile emission-peak intensity proxy rather than a discrete peak identified in every complex profile. We use the k2 intensity proxy as a qualitative diagnostic of the chromospheric thermal response, rather than as a temperature at a fixed geometrical height \citep{leenaarts_formation_2013}. Its regional sum also depends on the emitting area. Observed k profiles at the precursor site are shown in Figure~\ref{fig:supp_profiles} (Appendix~\ref{app:profiles}).

The \MgII\ k3 ridge is the elongated bright structure in the k3 intensity-excess maps in Figure~\ref{fig:mg}(a,c). The black-backed gold curve marks its adopted axis. For the spatial Doppler analysis, the \emph{ridge region} contains mapped ROI pixels within a Euclidean distance of $2''$ from the frame-specific axis and within the available raster FOV; its boundaries are shown in solid yellow. The dashed teal boundaries enclose the broader upper ROI sampled by IRIS. This broader region includes the ridge region in the two displayed rasters and is not an independent quiet-background control.
The ridge is directly identifiable in rasters centered from approximately 19:11:11 to 19:19:29~UT. Its axis is specified in reference rasters centered near 19:12:22, 19:15:56, and 19:17:42~UT, interpolated between those times, and held at the earliest or latest reference geometry outside that interval. These manually selected axes define the sampling geometry rather than a reconstructed magnetic-field axis. The visible-ridge interval and the interval between reference axes are distinct. The cross-ridge Doppler analysis in Figure~\ref{fig:mg} is restricted to the visible-ridge interval. For the central-reversal time series, the adopted ridge region is additionally restricted to the upper ROI and retained as a sampling reference outside the visible interval; this does not imply continued detection of a ridge.

To quantify the central reversal, we use the central-minimum intensity relative to the adjacent emission peaks:
\[
R_h=\frac{I_{\mathrm{h3}}}{I_{\mathrm{h2}}},
\qquad
R_k=\frac{I_{\mathrm{k3}}}{I_{\mathrm{k2}}},
\]
where h3 and k3 are the adopted central-reversal intensities and h2 and k2 are the corresponding full-profile emission-peak intensity proxies.
For each raster, Figure~\ref{fig:timing}(c) shows the median of the per-profile values in the ridge region and in the broader upper ROI (Figure~\ref{fig:mg}(a,c)); the upper-ROI sample includes the ridge region, whose footprint follows the time-dependent reference axis.
Larger values of $R_h$ or $R_k$ indicate a shallower central reversal, whereas smaller values indicate a deeper reversal; the ratio is not a direct optical-depth measurement.
In Figure~\ref{fig:timing}(d), the k2 sum includes k2 intensity proxies above 40 data numbers (DN) within the IRIS-covered upper ROI at Solar $Y>375\arcsec$, excluding the photospheric footpoint. The k3 median is calculated over valid intensity samples in the full IRIS-covered upper ROI. Each signal is divided by its own value in the first raster (R00); the two curves are different spatial statistics, not identically constructed averages.

For each complete raster, the regional \MgII\ k3 redshift and blueshift signals in Figure~\ref{fig:timing}(b) are calculated over the IRIS-covered upper ROI. The redshift signal is
\[
S_{\rm red}=\sum_{v_{k3}>3~\kms}(v_{k3}-3~\kms),
\]
and the blueshift signal is
\[
S_{\rm blue}=\sum_{v_{k3}<-3~\kms}(-v_{k3}-3~\kms).
\]
These are sums of velocity excess beyond the adopted threshold, not emission intensities. Each series is normalized to its own R00 value, \(S_{\rm red}/S_{{\rm red},0}\) and \(S_{\rm blue}/S_{{\rm blue},0}\). This region-integrated timing diagnostic is distinct from the ridge-relative spatial analysis in Figure~\ref{fig:mg}.

To describe the spatial distribution of the k3 Doppler signal across the ridge, we use the signed cross-ridge offset $d_{\rm cr}$, measured perpendicular to the k3 ridge axis and positive toward increasing Solar X. This coordinate is independent of the \Ha\ coordinate $n$ defined above. Redshifted samples have $v_{k3}>+3~\kms$ and blueshifted samples have $v_{k3}<-3~\kms$. For each raster, the signed difference between their median positions is
\[
\Delta d_{\rm cr}={\rm median}(d_{{\rm cr},{\rm red}})-{\rm median}(d_{{\rm cr},{\rm blue}}).
\]
Its absolute value is the \emph{cross-ridge red--blue separation}. For each stage, we take the absolute difference between the stage medians of the per-raster redshifted and blueshifted positions, not the median of the individual-raster absolute separations. Permutation tests assess the per-raster positional separation, while raster bootstrap and leave-one-raster-out tests assess the stage-level uncertainty (Appendix~\ref{app:stats}).

The IRIS\(^2\) temperature context is the gas temperature of the best-matching representative atmosphere averaged over \(-5\leq\log_{10}\tau_{500}\leq-4\), where $\tau_{500}$ is the continuum optical depth at 500~nm:
\begin{equation}
T_{[-5,-4]}=\frac{1}{1~\mathrm{dex}}\int_{-5}^{-4}T(\log\tau_{500})\,d\log\tau_{500}.
\end{equation}
This one-dex optical-depth average provides auxiliary thermal context, not a temperature at a fixed geometrical height or the unique formation depth of the k2 peaks \citep{leenaarts_formation_2013}. The k2 intensity and \(T_{[-5,-4]}\) describe the same \MgII\ spectra and are not independent temperature measurements.

The observed Mg~\textsc{ii} k width is a second-moment full-width-at-half-maximum (FWHM)-equivalent calculated from positive continuum-subtracted intensities over \(\pm80~\kms\):
\begin{equation}
\sigma_v^2=\frac{\sum I_+(v)(v-\bar v)^2}{\sum I_+(v)},\qquad
W_k=2\sqrt{2\ln2}\,\sigma_v,
\end{equation}
where \(I_+\) retains only positive continuum-subtracted intensity in this fixed velocity window and $\bar v$ is its intensity-weighted mean velocity.
For self-reversed profiles, \(W_k\) is an observed profile-envelope metric rather than a literal half-maximum width or a nonthermal velocity.

For \SiIV, each regional time-series point is the sum over accepted pixels in the IRIS-covered upper ROI of the positive continuum-subtracted line signal integrated over velocity. Both the peak and integrated signal-to-noise ratios (S/N) must exceed 3. The unnormalized regional integral has units of $(\mathrm{DN\,s^{-1}})(\mathrm{km\,s^{-1}})$; each line is divided by its own R00 value in Figure~\ref{fig:timing}(e). This time-series product is distinct from the wavelength-integrated map quantity in Figure~\ref{fig:supp_multilayer}. To time the \SiIV\ rise, we use the rasters before the Stage~1 onset as a reference and define the event threshold as the reference median plus 3 times 1.4826 times the median absolute deviation. The rise begins with the first of two consecutive rasters above this threshold.

The multi-layer spatial evolution at AIA cadence is shown in Figure~\ref{fig:supp_multilayer} (Appendix~\ref{app:multilayer}).

\subsection{NIRIS magnetic field}
\label{sec:niris_data}

GST/NIRIS obtained full-Stokes spectropolarimetry of the photospheric Fe~\textsc{i} 15648.5~\AA\ line.
The signed LOS field \(B_{\rm LOS}\) was derived with the center-of-gravity method from the wavelength separation of the circularly polarized \(I+V\) and \(I-V\) absorption-profile centroids \citep{rees_line_1979,uitenbroek_accuracy_2003}.
The sequence spans 19:06:47--19:59:49~UT with a median cadence of 51.0~s and was stabilized and registered to the common Earth-view frame.
Within the broader photospheric footpoint region, we use a fixed compact subregion containing the dominant negative-polarity patch to measure positive, negative, and unsigned fluxes, thresholded area, component centroids, and morphology above \(3\sigma\simeq21.95\)~G. The dashed boundary in Figure~\ref{fig:context}(e) outlines this subregion for statistical analysis and is distinct from the solid negative-field threshold contour.
The first NIRIS frame is excluded because of seeing instability.
Throughout the paper, \emph{footpoint} denotes the photospheric magnetic region identified in NIRIS.
The stereoscopic measurements locate EUV-emitting features but are not interpreted as direct measurements of the photospheric magnetic footpoint.

\subsection{\HRIEUV--AIA stereoscopy}
\label{sec:stereo_data}

We identify corresponding EUV features in paired \HRIEUV\ and AIA images. A selected position in one view constrains the possible counterpart to an epipolar line in the other view. The paired image coordinates define a stereoscopic tie point, and the closest approach of the two viewing rays gives an estimate of its three-dimensional position. The reconstructed positions are shown as dots in Figure~\ref{fig:stereo}(d,e).

We obtained 39 closest-ray stereoscopic solutions in seven frame pairs at 19:08:47, 19:10:22, 19:11:22, 19:11:47, 19:12:32, 19:15:32, and 19:17:57~UT, denoted Epochs~E1--E7.
Two near-photospheric Epoch~E5 locations were constructed by visually extending the observed EUV leg and are excluded from the retained observational sample used for the scientific interpretation.
The retained observational sample therefore contains 37 directly identifiable tie points.
A closest-ray solution gives their three-dimensional positions and radial heights above the adopted photosphere.
For each point, 500 Monte Carlo realizations independently perturb both image coordinates in each view with Gaussian $1\sigma=1$ pixel errors. The resulting single-point height uncertainties are typically 0.63--0.65~Mm and quantify the imposed pixel-coordinate perturbations; they do not include systematic uncertainties from manual feature association or cross-instrument registration.
The main-text stereoscopic analysis focuses on the recurrent compact kernel and selected Stage~1 and Stage~2 bright strands, while all 37 retained points are displayed in Figures~\ref{fig:supp_stereo_a} and \ref{fig:supp_stereo_b} (Appendix~\ref{app:stereo}).

\subsection{Registered AIA DEM diagnostics}
\label{sec:dem_data}

The registered AIA DEM sequence combines the 94, 131, 171, 193, 211, and 335~\AA\ channels on a fixed world-coordinate-system grid. It contains 125 time steps from 19:05:10 to 19:29:58~UT at a median cadence of 12.0~s and covers \(\log_{10}T\,[\mathrm{K}]=5.7\)--7.1. The DEMs were recovered with the regularized inversion method of \citet{Hannah2012}.
We define the total emission measure (EM) as
\begin{equation}
{\rm EM}=\int {\rm DEM}(T)\,dT
\end{equation}
and integrate three temperature intervals: cool, \(5.7\leq\log_{10}T<6.0\); warm, \(6.0\leq\log_{10}T<6.3\); and hot, \(\log_{10}T\geq6.3\). We also calculate the emission-measure-weighted logarithmic temperature,
\begin{equation}
\left\langle\log_{10}T\right\rangle_{\rm EM}
=
\frac{\int {\rm DEM}(T)\log_{10}T\,dT}
{\int {\rm DEM}(T)\,dT}.
\end{equation}
The temperature-band EM curves characterize the plasma response in different temperature ranges, not separate geometrical layers. For the light curves in Figure~\ref{fig:timing}(f), the EM is averaged over valid pixels in the fixed AIA event region outlined in Figure~\ref{fig:dem} (180~arcsec$^2$, or $9.5\times10^{17}$~cm$^2$), and each curve is divided by its median over 19:05--19:06~UT. The density and estimated event-associated thermal content are derived from the common-baseline EM increase in Section~\ref{sec:geometry_results}, under the assumptions stated there.

\section{Results}
\label{sec:results}

\subsection{Pre-event structure and a compact multithermal precursor}
\label{sec:precursor_results}

Before the first EUV brightening, the \Ha\ line-center images show a curved, flux-rope-like thread embedded among straighter fibrils (Figure~\ref{fig:halpha}(a)). The thread is visibly misaligned with the surrounding fibrils by approximately $10^\circ$--$20^\circ$ in the plane of the sky, and this projected misalignment persists through both releases. Beneath this structure, the photospheric footpoint region is dominated by negative polarity (Figure~\ref{fig:context}(e)). Within the compact NIRIS statistical subregion, the negative-flux magnitude increases by about \(10\%\) during the event and the thresholded magnetic patches coalesce; no opposite-polarity signal or cancellation is detected above the adopted \(3\sigma\) threshold. The magnetic-patch centroids show non-monotonic motions rather than sustained rotation in one direction.

The lower atmosphere within this structure becomes active several minutes before the EUV event. A weak k2 enhancement is already present in the EUV source region near the later recurrent kernel. At the selected source position it is sampled at approximately 19:04:54 and 19:05:30~UT, 3.8 and 3.2~min before the Stage~1 onset. This compact brightening is the only conspicuous k2 enhancement within the upper ROI in R02 (Figure~\ref{fig:supp_multilayer}, row~(c)); we call it the \emph{compact precursor source}. Its footprint is shown in Figure~\ref{fig:supp_profiles}(a). Figure~\ref{fig:timing} provides the regional temporal context, while Figures~\ref{fig:supp_multilayer} and \ref{fig:supp_profiles} show the source location and observed spectral profiles.

A brief, stronger multithermal precursor pulse is sampled at the same source in R02, with source-pixel exposure times of approximately 19:06:01--19:06:05~UT, 2.6--2.7~min before Stage~1. During this pulse, the k2 intensity increases, the best-matching IRIS\(^2\) atmosphere has a higher \(T_{[-5,-4]}\), the \MgII\ k profile broadens, opposite-signed k3 velocities appear, and a compact \SiIV\ 1394~\AA\ signal becomes detectable. The precursor and the later recurrent EUV kernel lie within the same EUV source region, with a projected centroid separation of approximately \(1.9\arcsec\).

The observed width increases from about 43--45~$\kms$ before the pulse to about 48~$\kms$ during it, then decreases. The corresponding median $W_k$ values are 42.7 and 44.6~$\kms$ at approximately 19:04:54 and 19:05:30~UT, 47.5~$\kms$ during the pulse, and 45.2~$\kms$ at approximately 19:06:41~UT. The compact \SiIV\ signal and the largest width change are short lived, whereas the k2 enhancement remains visible at the latter time. Subsequently, the compact k2 maximum weakens while enhanced emission extends along the local fibril direction. The lower atmosphere thus evolves from persistent low-level activity, through a brief multithermal pulse, to a more extended k2 brightening before the first EUV release.

\subsection{Morphological and Doppler reorganization across the two releases}
\label{sec:stage1_results}

The \HRIEUV\ light curve separates the event into two brightenings from the same EUV source region (Figure~\ref{fig:eui}).
During Stage~1, a compact kernel repeatedly brightens and multiple resolved strands are illuminated outward from the source.
The strands acquire a helical-looking, entangled appearance and fill a widening channel.
Stage~2 later illuminates fewer, straighter strands inside the same channel.

\paperanimation{Movie1_Figure2_EUI.mp4}{\figtwofile}{0.99}
{\HRIEUV\ 174~\AA\ morphology of the two-stage event.
(a) Full \HRIEUV\ image showing the teal-outlined strip used for the cutout sequence and source-region photometry.
(b)--(h) Rotated cutouts with the apparent jet axis oriented approximately vertically.
Panels (b)--(h) show selected times from before Stage~1 to the end of Stage~2.
(i) Native-cadence source-region intensity, divided by its median over 19:05--19:06~UT (as in Figure~\ref{fig:timing}(a)), with the displayed times and the Stage~1 and Stage~2 intervals.}
{fig:eui}

The curved \Ha\ thread identified before the event takes part in both releases (Figure~\ref{fig:halpha}).
The upper ROI reaches its sampled \Ha\ blueward-shift maximum at 19:12:13~UT, within one \Ha\ sampling interval of the Stage~1 EUV peak; the sequential wavelength sampling does not resolve a physical lag.

The time--distance maps in Figure~\ref{fig:halpha}(e--g) show selected intensity features and abrupt changes during both stages.
Track A in Slit A has an apparent pattern speed of $+2.20\pm0.24~\kms$ toward increasing $n$ from 19:10:08 to 19:15:07~UT during Stage~1.
Track B in Slit B begins late in Stage~1 and continues through Stage~2 with an apparent pattern speed of $+1.47\pm0.21~\kms$ toward increasing $n$, and Track B$'$ in the same slit is a tentative counter-directed feature during Stage~2.
Track C in Slit C has an apparent pattern speed of $+3.06\pm0.55~\kms$ toward increasing $s$.
Feature appearances, disappearances, and abrupt slope changes mark morphological changes that occur between consecutive images. These changes are concentrated within the amber reconfiguration region in Figure~\ref{fig:halpha}(a--c), whose extent along each slit is indicated in panels (e--g).

The \Ha\ line-center and wing images both change in step with Stage~1.
During Stage~2, the line-center images again show clear structural changes and apparent motions, whereas the wing evolution is more gradual and lacks an equally clear correspondence with the second release.

The optical-flow diagnostics in Figure~\ref{fig:halpha}(d) show normalized residual image changes in the thread-core region about twice those in the quiet-control region during both stages. The median normalized residual is 0.192 in the thread core in each stage, compared with 0.084 and 0.104 in the quiet control for Stages~1 and~2, respectively. These residuals indicate image changes not accounted for by the inferred apparent displacement and are consistent with the appearance, disappearance, and intensity evolution of features in the \Ha\ image sequence.

\paperanimation{Movie2_Figure3_Halpha.mp4}{\figthreefile}{0.99}
{H$\alpha$ evolution and localized morphological changes. (a)--(c) Line-center images before Stage~1, during Stage~1, and during Stage~2. The cyan curve marks the adopted \Ha\ axis, and the arrows indicate the positive $s$ and $n$ directions. In (a), white solid and purple dashed boundaries identify the thread-core and quiet-control regions used for the optical-flow statistics in (d). The reconfiguration region has an amber dotted boundary in (a) and amber shading in (b,c); it is the same region outlined in Figure~\ref{fig:context}(d). Orange, magenta, and green dashed paths mark Slits A, B, and C. (d) Optical-flow post-remapping residual RMS and correlation gain. (e)--(g) Time--distance maps along Slits A--C. Cyan lines show feature tracks (Track B$'$ is tentative), and arrows mark where features appear, turn, or change slope; they do not identify a continuously tracked plasma parcel. The horizontal amber bands mark the extent of the reconfiguration region along each slit, not the duration of activity. The strips at the top mark the two stages. (h) Independently normalized \HRIEUV\ 174~\AA\ emission and the \Ha\ blueward-shift proxy.}
{fig:halpha}

In the IRIS spectra, the releases are accompanied by an elongated \MgII\ k3 bright ridge (Figure~\ref{fig:mg}(a,c)), identifiable in rasters centered from approximately 19:11:11 to 19:19:29~UT. Redshifted and blueshifted k3 samples occupy opposite sides of this ridge. Their stage-level cross-ridge separation decreases from \(1.47\arcsec\) in Stage~1 to \(0.89\arcsec\) in Stage~2 (Figure~\ref{fig:mg}(e)). All seven Stage~1 rasters, centered from approximately 19:11:11 to 19:14:44~UT, retain the same opposite-side ordering and have permutation $p<0.05$. The six Stage~2 rasters, centered from approximately 19:16:31 to 19:19:29~UT, retain that ordering, with three having $p<0.05$. The stage-level positions and bootstrap intervals are given in Appendix~\ref{app:stats}.

Most of the nominal decrease in separation comes from the redshifted median moving toward the ridge, whereas the blueshifted median changes comparatively little. The change therefore describes an asymmetric redistribution of the Doppler-selected samples rather than a symmetric contraction of a rotating tube.

Repeating the analysis at \(\lvert v\rvert>5~\kms\) gives separations of \(1.78\arcsec\) in Stage~1 and \(1.20\arcsec\) in Stage~2, preserving the same stage ordering.
The \(\lvert v\rvert>10~\kms\) sample is too sparse for a robust comparison.
Bootstrap resampling of rasters and leave-one-raster-out calculations also retain the larger Stage~1 separation, so the contrast is not driven by one exceptional raster. The larger Stage~1 separation is also retained when the k3 velocity zero point is shifted by up to \(\pm5~\kms\) (Appendix~\ref{app:robust}).
The two rasters between the stages, centered at approximately 19:15:19 and 19:15:56~UT, still show significant opposite-side organization, so the pattern does not disappear immediately after Stage~1.

The decrease in the cross-ridge separation parallels the change from the multithreaded, helical-looking, and widening Stage~1 EUV emission to fewer and straighter Stage~2 strands.
The two releases therefore differ jointly in their visible strand morphology and in the organization of the chromospheric Doppler signal, although both originate in the same EUV source region and involve the same \Ha\ thread.

\paperfigure{\figfourfile}{0.99}
{\MgII\ k3 ridge and cross-ridge Doppler organization. (a), (b) Pre-event-subtracted k3 intensity excess and k3 velocity for the Stage~1 raster centered at 19:12:22~UT. (c), (d) Corresponding maps for the Stage~2 raster centered at 19:17:42~UT. Black-backed gold curves mark the adopted ridge axes, and the color bars give common scales for the two rasters. In (a) and (c), solid yellow boundaries enclose mapped ROI pixels within $2\arcsec$ of the frame-specific ridge axis; dashed teal boundaries enclose the broader upper ROI within the IRIS FOV. The regions overlap. Blue and red contours mark $-3$ and $+3~\kms$, respectively; positive velocity denotes redshift. (e) Median signed cross-ridge positions of redshifted samples (red circles) and blueshifted samples (blue squares) during the visible-ridge interval, R11--R25. Positive cross-ridge offset is toward increasing Solar X. Filled and open symbols denote permutation $p<0.05$ and $p\geq0.05$, respectively. Points are placed at raster midpoints, with horizontal spans showing acquisition intervals, not positional uncertainties. Amber, gray, and teal shading indicate Stage~1, the transition interval, and Stage~2. The stage-level separations are $1.47\arcsec$ and $0.89\arcsec$: each is the absolute difference between the stage medians of the red and blue positions, not the median of the individual-raster absolute separations.}
{fig:mg}

\subsection{Persistent spectroscopic response and the Stage~2 release}
\label{sec:stage2_results}

The \MgII\ spectra and \SiIV\ emission continue to evolve after the first EUV peak and into the second release, showing a persistent spectroscopic response across the two stages (Figure~\ref{fig:timing}).

For the pre-event comparison, we use the mean and sample standard deviation $\sigma$ of R00, R01, R03, R04, R05, and R06, excluding the precursor raster R02. These six rasters define the mean $\pm3\sigma$ bands in Figure~\ref{fig:timing}(c,e); $\sigma$ describes pre-event scan-to-scan variability, not a confidence interval. The baseline values and the last complete raster before Stage~2 are listed in Table~\ref{tab:pre_event_r19}.

Figure~\ref{fig:timing}(c) shows the Mg~\textsc{ii} h and k central-reversal ratios defined in Section~\ref{sec:iris_data}. Their median values range from approximately 0.33 to 0.60, with larger values indicating shallower reversals.

Across the broader upper ROI, the central-reversal depths are comparatively stable before Stage~1. A modest reduction in depth begins in the raster centered near 19:11:11~UT, and the reversals remain slightly shallower, by less than about $2\sigma$ of the pre-event variability, through the inter-stage interval. At the beginning of Stage~2 they first deepen, then become briefly shallower near the second EUV peak, with the reduced depth persisting through approximately 19:18:53~UT. Subsequent rasters show renewed deepening.

The ridge region shows a larger reduction in central-reversal depth than the broader upper ROI. The h and k reversals become shallowest at approximately 19:13:33~UT, about 1.2~min after the Stage~1 EUV peak, and then deepen gradually. A brief reduction in depth near the Stage~2 peak is followed by renewed deepening. Thus, the strongest change is concentrated around the ridge during Stage~1, whereas the second peak produces a shorter-lived change. The same sequence is retained when the sampling is restricted to the 452 spatial pixels common to all adopted ridge regions (Appendix~\ref{app:robust}), so it is not produced solely by the changing geometric footprint.

A same-pixel comparison with the pre-event spectra shows that the shallower central reversal during the Stage~1 decay is associated mainly with brightening at the central minimum rather than dimming of the emission peaks. The paired logarithmic intensity changes correspond to increases of about 42\% and 47\% at h3 and k3, respectively, compared with only about 6\%--7\% in the h2 and k2 peak-intensity proxies. The central reversal subsequently deepens in both the ridge region and the broader upper ROI, although in the ridge region it remains shallower than before the event until after Stage~2.

A thread-shaped \SiIV\ enhancement becomes clearly visible during the Stage~1 decay and develops along the channel. The 1394~\AA\ rise begins in the raster centered near 19:12:58~UT (19:12:41--19:13:14~UT), about 0.6~min after the Stage~1 EUV peak. The regional signal continues to build between the releases and reaches its event-time maximum in the raster centered near 19:17:42~UT, coincident with the Stage~2 EUV peak at the IRIS cadence. It returns below the event threshold by approximately 19:22:44~UT. The 1402~\AA\ line follows the same overall evolution.

R19, centered at 19:15:56~UT and acquired over 19:15:39--19:16:12~UT, is the last complete raster before Stage~2 begins at 19:16:31~UT. In this raster, the regional \SiIV\ 1394 and 1402~\AA\ signals exceed their pre-event means by about $21\sigma$ and $19\sigma$, the regional k3 blueshift signal by about $14\sigma$, and the ridge-region h3/h2 and k3/k2 ratios by about $5\sigma$. The k3 redshift signal and the central-reversal ratios in the broader upper ROI are within $1\sigma$ of their pre-event means. Thus, the second release begins while the regional \SiIV\ and k3 blueshift signals remain enhanced and the ridge-region central reversals remain shallower. The persistent reversal change is more pronounced in the ridge sample than in the broader, overlapping sample; this comparison does not establish that the surrounding atmosphere as a whole has recovered.

Together, Figures~\ref{fig:halpha}--\ref{fig:timing} show that H$\alpha$ morphological changes accompany both EUV releases while the \MgII\ central reversal and \SiIV\ intensity continue to evolve between them.

During the Stage~2 decay, the thresholded k2 sum increases, while the k3 median over the full IRIS-covered upper ROI varies more modestly relative to its own R00 value (Figure~\ref{fig:timing}(d)). Compared with the compact precursor sampled around 19:06~UT, the peak local k2 intensity is about 6\% lower (ratio 0.943), whereas the area above 40~DN is nearly four times as large (3.91) and the thresholded sum is about three times as large (3.13). The larger regional k2 response is therefore associated mainly with a larger emitting area, not a higher peak local intensity. Both k2V and k2R increase over the well-sampled decay interval represented by raster midpoints from approximately 19:18:17 to 19:21:15~UT, with a larger change in k2V, while a weaker positive k3 response accompanies much of the interval. Figure~\ref{fig:supp_multilayer} shows the expansion of the k2-emitting area.  

\paperfigure{\figfivefile}{0.82}
{Multi-layer temporal evolution. (a) \HRIEUV\ 174~\AA\ intensity summed over the teal strip in Figure~\ref{fig:eui}(a), divided by its median over 19:05--19:06~UT. (b) Regional \MgII\ k3 Doppler signals in the IRIS-covered upper ROI: sums of redshift and blueshift velocity excess beyond $3~\kms$, each divided by its own R00 value. Red and blue denote Doppler sign, not emission intensity. (c) Median per-profile central-reversal ratios, $I_3/I_2$, for h and k in the ridge region and broader upper ROI. Larger ratios indicate shallower reversals. The regions overlap; the broader ROI is not an independent quiet control. Filled ridge markers identify R11--R25, when the ridge is visible; open markers use the reference sampling geometry outside that interval. (d) The k2 sum (teal) includes intensity proxies above 40~DN within the IRIS-covered upper ROI at Solar $Y>375\arcsec$; the k3 median (orange-red) uses the full IRIS-covered upper ROI. Each is divided by its own R00 value; these are different spatial statistics, and their colors do not denote Doppler sign. (e) Regional \SiIV\ 1394 and 1402~\AA\ velocity-integrated line signals, each divided by its own R00 value. (f) AIA cool, warm, and hot EM, each divided by its median over 19:05--19:06~UT; normalized amplitudes do not measure the relative absolute EM of the bands. Amber and teal shading mark the two stages. Gray dashed lines in (b), (c), and (e) mark the R19 midpoint, 19:15:56~UT; its full scan, 19:15:39--19:16:12~UT, ends before Stage~2 begins at 19:16:31~UT. Horizontal segments denote scan acquisition intervals, not measurement errors. Horizontal bands in (c) and (e) show each curve's pre-event mean $\pm3$ sample standard deviations from R00--R06 excluding R02, in the curve's displayed normalization; they describe scan-to-scan variability, not confidence intervals. This figure gives regional temporal evolution, while Figures~\ref{fig:supp_multilayer} and \ref{fig:supp_profiles} provide the precursor's spatial and spectral context. The \Ha\ blueward-shift proxy is shown in Figure~\ref{fig:halpha}(h).}
{fig:timing}

\subsection{Height and thermal content of the EUV emission}
\label{sec:geometry_results}

The observations above associate both releases with the same EUV source region and evolving \Ha\ structure. Stereoscopy places the sampled EUV-emitting features at low heights (Figure~\ref{fig:stereo}).
The 37 retained stereoscopic tie points span \(0.58\)--\(4.17~\Mm\) above the photosphere, with typical statistical uncertainties of \(0.64~\Mm\).
A compact EUV kernel is measured at Epochs~E1--E5 and again at Epoch~E7.
These measurements cluster near \(2.2~\Mm\) at a common projected location and show no systematic rise.

The selected Stage~1 and Stage~2 bright strands outline families of inclined, locally arched structures.
Along representative bright lines, the radial height first increases and then decreases while the projected position advances along the apparent ejection direction.
At Epoch~E7, the selected points trace a left strand, a short central strand, and a right strand, and at Epoch~E5, three selected locations sample part of the Stage~1 right strand.
The sampled features thus outline low-lying, inclined, locally arched emitting structures a few Mm above the photosphere, rather than a vertically extended visible spire.

\paperfigure{\figsixfile}{0.99}
{Stereoscopic geometry of the two releases. (a), (b) Corresponding compact EUV features selected in the \HRIEUV\ 174~\AA\ and AIA 171~\AA\ images. The line in (b) is the epipolar constraint on the AIA counterpart. (c) The recurrent-kernel location projected onto the co-aligned \Ha\ image. (d), (e) Reconstructed positions of selected EUV-emitting features (tie points; colored dots) at representative Stage~1 and Stage~2 epochs. Dot colors indicate radial height above the adopted photosphere. K identifies the recurrent compact EUV kernel. The guide lines group features illuminated at the same observational epoch; they are not fitted plasma trajectories or reconstructed magnetic-field lines. The points are drawn from the retained sample in Figures~\ref{fig:supp_stereo_a} and \ref{fig:supp_stereo_b} and share their height color scale. (f) Radial heights of the recurrent compact EUV kernel at six independently measured epochs. The dashed horizontal line marks the median kernel height. Error bars show the statistical $1\sigma$ height uncertainty from the pixel-perturbation procedure in Section~\ref{sec:stereo_data}; registration and feature-association systematics are not included. All heights are measured above the adopted photosphere.}
{fig:stereo}

The AIA DEM diagnostics support an association of the low-lying EUV emission with plasma at coronal temperatures. The registered sequence shows two multithermal responses associated with the two \HRIEUV\ peaks (Figures~\ref{fig:timing}(f) and \ref{fig:dem}). The cool, warm, and hot EM bands increase during both stages and decrease between and after them. The peak region-averaged EM increments are similar: \(4.7\times10^{26}\) and \(4.4\times10^{26}~\mathrm{cm^{-5}}\) for Stages~1 and~2, respectively, relative to the common pre-event mean. Each increment is about 12 times the pre-event standard deviation, and about two-thirds is in the hot band (\(\log_{10}T\geq6.3\)).

We treat the EM excess above the common pre-event baseline as an event-associated emitting component to estimate its density and thermal content. Multiplying each region-averaged increment by the AIA event area gives a volume EM of approximately $4\times10^{44}~\mathrm{cm^{-3}}$. We adopt an LOS depth $L=1$--4~Mm and unit filling factor; this depth is an assumption, not a direct stereoscopic measurement of the LOS thickness. The resulting electron density is approximately $(1$--$2)\times10^9~\mathrm{cm^{-3}}$. With a characteristic temperature of 3~MK, the estimated thermal content $3n_e k_B T V$, where $V$ is the adopted emitting volume, is $(2.5$--$5)\times10^{26}$~erg at each peak. For fixed area and depth, a filling factor $f<1$ reduces this estimate by $f^{1/2}$.

Under the same geometric and filling-factor assumptions, the two peaks have comparable estimated event-associated thermal contents despite their different EUV morphologies and cross-ridge Doppler separations. These are order-of-magnitude estimates for an assumed excess-emission component, not measurements of the total energy released in either episode. Because both increments are measured relative to the same pre-event baseline, the Stage~2 estimate may include a residual contribution from Stage~1.

\paperanimation{Movie3_Figure7_DEM.mp4}{\figsevenfile}{0.99}
{AIA DEM-derived thermal diagnostics. The top row shows total EM and the bottom row the EM-weighted mean of $\log_{10}(T/\mathrm{K})$ at a pre-event time and the Stage~1 and Stage~2 peaks. The black-and-white outlines mark the AIA event region. All panels use the same registered coordinate grid and a common color scale within each row, with color bars on the right. These are total DEM-derived maps, not background-subtracted EM-increment maps; the thermal-content estimate in Section~\ref{sec:geometry_results} instead uses the region-averaged EM increment relative to a common pre-event baseline. Figure~\ref{fig:timing}(f) shows the EM time profiles, normalized by the median of each temperature band over 19:05--19:06~UT.}
{fig:dem}

\section{Discussion}
\label{sec:discussion}

The observations in Section~\ref{sec:results} support a picture of successive ejections within one evolving low-lying structure. A curved \Ha\ thread is present before the event; chromospheric activity precedes the first EUV brightening; the sampled EUV features lie a few Mm above the photosphere; and the chromospheric and TR spectra continue to evolve between the brightenings. We first discuss the emitting geometry and its association with the precursor (Section~\ref{sec:precursor_discussion}), then a component-reconnection interpretation (Section~\ref{sec:component_reconnection}), and finally the persistent spectroscopic response across the two releases (Section~\ref{sec:stratified_discussion}).

\subsection{Low-lying EUV emission and its chromospheric precursor}
\label{sec:precursor_discussion}

The compact k2 precursor shows that the first prominent EUV brightening does not mark the beginning of lower-atmospheric activity. Its brief chromospheric/TR pulse, with broadened profiles, opposite-signed k3 velocities, and compact \SiIV\ emission, is consistent with localized energy release preceding the main EUV brightening.
Small-scale IRIS reconnection events have likewise been detected first through bidirectional Mg~\textsc{ii}, C~\textsc{ii}, and \SiIV\ spectral components, followed by overlying multi-channel EUV brightening \citep{li_spectroscopic_2018}.
Other chromospheric jets progress from a weaker cool \Ha\ phase to a hotter EUV release as the dissipation region lengthens and reconnection intensifies \citep{tang_photospheric_2025}.

Stereoscopy locates the sampled EUV-emitting features within a few Mm of the photosphere, along inclined, locally arched structures (Figure~\ref{fig:stereo}(d,e)). The AIA DEM diagnostics indicate plasma at coronal temperatures in the associated event region. The precursor is close to the later recurrent kernel in projected position, but it has no stereoscopic height measurement. The kernel near 2.2~Mm may mark the vicinity of repeated energy release; its measured height does not directly locate the reconnection site.

Together, the low-lying EUV geometry and its spatial and temporal association with the \Ha, \MgII, and \SiIV\ activity are consistent with energy release within or near an evolving lower-atmosphere structure. The emitting geometry permits field-aligned motion and sequential illumination of low-lying strands, as in other low-lying brightenings and redirected jets \citep{zhukov_stereoscopy_2021,chen_transient_2021,yang_observational_2023}. This provides an alternative to interpreting the visible feature solely as plasma ejected along an extended coronal spire. Some jet-like features in high-resolution EUV images may therefore represent a coronal-temperature response within low-lying structures. The observed geometry does not determine the complete magnetic connectivity or the fate of all plasma, so it does not by itself establish or exclude a contribution to the solar-wind mass supply.

\subsection{A component-reconnection interpretation}
\label{sec:component_reconnection}

The combined \Ha\ and IRIS observations show that a curved cool structure is present before the first prominent EUV release and remains dynamically involved during both releases. Its approximately $10^\circ$--$20^\circ$ projected misalignment with the surrounding fibrils motivates consideration of an interaction between the curved structure and the surrounding field. We therefore consider component reconnection as a possible explanation for both releases within one evolving low-lying system. In this geometry, an azimuthal component of the flux-rope-like structure can oppose part of the ambient field while a finite guide-field component remains. The projected morphology does not uniquely determine the three-dimensional field directions.

The magnetic association adopted here differs from that discussed by \citet{lee_fine_2025}, who reported mixed polarity near the source. In our selected compact NIRIS subregion, no opposite-polarity signal or cancellation is detected above the adopted threshold. This non-detection is limited by the NIRIS sensitivity and the adopted subregion. Associating the emitting structures with particular photospheric patches relies in part on projected \Ha\ trajectories and is subject to uncertainty in feature identification and registration. We do not use this difference to exclude alternative magnetic configurations. Component reconnection describes the local interaction geometry and does not by itself establish or exclude interchange connectivity.

The Stage~1 strand morphology and cross-ridge Doppler pattern provide the observational context for this interpretation. Reconnection involving part of the twisted structure could transfer magnetic stress and twist into newly connected strands, which would then be heated and appear as the multithreaded, helical-looking EUV emission. Observed interactions between filament threads and neighboring fibrils or loops provide relevant examples of connectivity change and untwisting \citep{xue_observing_2016,kumar_formation_2015,li_magnetic_2016}. Opposite-sided Mg~\textsc{ii} Doppler signatures combined with resolved EUV rotational morphology have also been used to interpret untwisting in a larger jet \citep{panesar_buildup_2025}.

The \Ha\ thread and its spectral environment continue to evolve after Stage~1. One possible interpretation is that Stage~1 only partially reconfigures the magnetic structure, allowing remaining or adjacent flux to interact in the same or a nearby region during Stage~2. The straighter Stage~2 strands and smaller cross-ridge separation are compatible with partial redistribution of twist, but the separation is not a measure of rotation speed or magnetic twist. Field-aligned counterflows, reconnection outflows, and changes in which strands emit can also contribute. The persistent spectroscopic response does not by itself measure the remaining magnetic free energy or identify the Stage~2 trigger.

\subsection{Atmospheric response across successive releases}
\label{sec:stratified_discussion}

The two releases differ in their EUV strand morphology and cross-ridge Doppler distribution, but their estimated event-associated thermal contents are of the same order under the adopted assumptions (Section~\ref{sec:geometry_results}). Stage~2 also coincides with the principal event-time \SiIV\ peak (Section~\ref{sec:stage2_results}). The simpler Stage~2 morphology and smaller Doppler separation are therefore not accompanied by a proportionally weaker observed thermal response. This comparison does not establish equal total energy release in the two episodes.

The evolving \MgII\ central reversal records changes in the chromospheric spectral state. Changes in temperature, density, source-function coupling, opacity, and unresolved velocities can affect the central reversal \citep{carlsson_what_2015,rubio_da_costa_parameter_2017}; its persistence alone does not determine the duration of heating.
\SiIV\ samples plasma at characteristic temperatures of order $10^5$~K under ionization equilibrium \citep{depontieu_interface_2014}, and its intensity is not a direct measure of the heating rate. Its emission continues to build toward the second release, when the \MgII\ central reversal briefly becomes shallower again. Changes in the heating history, density, emitting volume, and cooling through \SiIV-emitting conditions can all contribute, and the line intensities do not uniquely separate these possibilities. Similarly, because the regional k2 sum during the Stage~2 decay increases mainly through a larger emitting area, it does not indicate a corresponding increase in local temperature. A decline in the regional k3 blueshift signal likewise does not by itself establish a downflow.

In the last complete raster before Stage~2 begins, the regional \SiIV\ emission and k3 blueshift signal remain enhanced, while the ridge-region \MgII\ central reversal is still shallower than before the event (Section~\ref{sec:stage2_results}). Together, these measurements show a persistent spectroscopic response between the two ejections. The central-reversal change is more pronounced in the ridge than in the broader, overlapping sample. Thus, the second ejection begins before the atmosphere has returned to its pre-event spectroscopic state. Its spectral signatures may depend on the atmospheric changes associated with the first ejection as well as on the new energy input. Models of successive ejections should therefore follow the atmospheric evolution from the first event into the second, rather than treat both as energy inputs into the same undisturbed atmosphere.

\section{Summary and Conclusions}
\label{sec:summary}
Although the second ejection has straighter EUV strands and a smaller cross-ridge separation of redshifted and blueshifted \MgII\ k3 samples, the two peaks have comparable EM increments and estimated thermal contents under the same assumptions. Before the second ejection begins, the regional \SiIV\ emission and \MgII\ k3 blueshift signal remain enhanced, and the \MgII\ central reversal in the ridge remains shallower than before the event. The second ejection therefore occurs before the atmosphere has returned to its pre-event spectroscopic state. Its spectral signatures may depend on the atmospheric changes associated with the first ejection as well as on the subsequent energy input.

Building on the earlier EUI/GST study of this event and its energetics \citep{lee_fine_2025}, we use IRIS spectroscopy, \HRIEUV--AIA stereoscopy, and AIA DEM diagnostics to follow the pre-event atmosphere, the emitting geometry, and the evolution between and after the releases. The observational findings supporting this interpretation are as follows:

\begin{enumerate}
\item
A curved \Ha\ thread is present before the first EUV brightening. Weak \MgII\ k2 emission at a compact source is detected 3.2--3.8~min before the onset of the first ejection (Stage~1). Later, 2.6--2.7~min before the same onset, this source shows a brief enhancement in both chromospheric \MgII\ k2 and transition-region \SiIV\ emission. This multithermal precursor is accompanied by broader \MgII\ k profiles and both redshifted and blueshifted k3 samples.

\item
The sampled EUV emission lies at approximately 0.6--4.2~Mm above the photosphere along inclined, locally arched structures, with the recurrent compact kernel near 2.2~Mm. The associated DEM response indicates plasma at coronal temperatures within this low-lying emitting system.

\item
Stage~1 shows multithreaded, helical-looking EUV strands, whereas Stage~2 illuminates fewer, straighter strands. The cross-ridge separation of redshifted and blueshifted k3 samples decreases from $1.47\arcsec$ to $0.89\arcsec$, mainly through a change in the redshifted positions. Despite this morphological and Doppler difference, the peaks have comparable EM increments and estimated event-associated thermal contents of a few $10^{26}$~erg under the same assumptions.

\item
In the last complete IRIS raster before Stage~2, the regional \SiIV\ and k3 blueshift signals remain enhanced relative to their pre-event levels, while the ridge-region \MgII\ central reversal remains shallower. Thus, the chromospheric and TR spectra have not all returned to their pre-event states when the second ejection begins.
\end{enumerate}

Component reconnection between the curved structure and the surrounding field, with partial magnetic reconfiguration during Stage~1, provides a possible interpretation of the magnetic evolution across the two releases. Models of successive small-scale releases should reproduce the chromospheric and TR evolution between the brightenings, as well as the EUV brightenings themselves.

\section*{Code Availability}
Reusable implementations of the SolO/EUI \HRIEUV\ residual-jitter correction, \HRIEUV--FSI pointing registration, and \HRIEUV--AIA stereoscopic tie-pointing workflow are publicly available at \url{https://github.com/ny-huang/solo-eui-stereo-toolkit} under the MIT License.
The repository contains portable Python and IDL/SolarSoft routines, documentation, calibration records, and a worked 2022 October 29 example.

\facilities{SolO (EUI), IRIS, SDO (AIA), BBSO/GST (NIRIS, VIS)}

\software{IDL, SolarSoft \citep{Freeland2012}, Astropy, SunPy, aiapy, demreg, NumPy, SciPy, scikit-image, OpenCV, Matplotlib}

\clearpage
\appendix
\setcounter{figure}{0}
\renewcommand{\thefigure}{A\arabic{figure}}
\renewcommand{\theHfigure}{A\arabic{figure}}

\section{Supplementary Observational Details}
\label{sec:supplement}

\subsection{Image stabilization and cross-instrument registration}
\label{app:align}

The \HRIEUV\ sequence was corrected for residual frame-to-frame jitter and anchored to an absolute \HRIEUV--FSI pointing reference.
The GST \Ha\ and NIRIS sequences received independent temporal-translation corrections.
Figure~\ref{fig:supp_alignment} summarizes the corrections and representative registration checks.

\paperanimation{MovieS1_FigureS1_EUI_stabilization.mp4}{\figsalignfile}{0.99}
{Image stabilization and registration checks.
(a) Representative \HRIEUV\ image with the morphology-strip outline (white) and the sampling line for (b), (c). Panels (b), (c) show the same time--distance sampling after and before stabilization, respectively. Panel (d) shows the residual adjacent-frame displacement before and after correction. The median displacement decreases from 0.524 to 0.046 pixels.
The temporal corrections preserve the inherited absolute registration convention.}
{fig:supp_alignment}

\subsection{Stereoscopic tie points}
\label{app:stereo}

The stereoscopic analysis yielded 39 closest-ray solutions in seven epochs.
The retained observational sample comprises 37 directly identifiable points; two visually extended, near-photospheric Epoch~E5 locations excluded from the scientific sample are not displayed.
Figure~\ref{fig:supp_stereo_a} covers Epochs~E1--E4, and Figure~\ref{fig:supp_stereo_b} covers Epochs~E5--E7.
Each row shows \HRIEUV\ 174~\AA, AIA 171~\AA\ with the point-specific epipolar loci, and the co-aligned H\(\alpha\) centroid-velocity proxy.
The point colors encode radial height on the same scale as the height-coded points in Figure~\ref{fig:stereo}.
Together, the figures show the recurrent compact kernel, directly visible lower endpoints, and Stage~1 and Stage~2 bright-strand sampling that supports the low-lying, inclined, and locally arched emitting geometry.
The points are independent epoch-by-epoch measurements and do not represent one continuously tracked plasma element or reconstructed magnetic field line.

\paperfigure{\figsstereoafile}{0.88}
{Retained stereoscopic tie points for Epochs~E1--E4. K identifies the recurrent compact EUV kernel.
Each row corresponds to one epoch; the columns show \HRIEUV\ 174~\AA, AIA 171~\AA\ with the point-specific epipolar loci, and the co-aligned \Ha\ centroid-velocity proxy map.
All retained stereoscopic tie points for each epoch are shown, with color indicating radial height above the adopted photosphere on the same 0.58--4.17~Mm scale as Figure~\ref{fig:stereo}.
Light-gray canvas labeled ``outside H-alpha coverage'' identifies projected locations beyond the available \Ha\ field rather than zero-valued Doppler data.
The panels show the spatial distribution of recurrent-kernel, directly visible lower-endpoint, and bright-strand measurements; they do not trace one continuously tracked plasma parcel.
All heights are radial above the adopted photosphere.}
{fig:supp_stereo_a}

\paperfigure{\figsstereobfile}{0.99}
{Retained stereoscopic tie points for Epochs~E5--E7. K identifies the recurrent compact EUV kernel.
Each row corresponds to one epoch; the columns show \HRIEUV\ 174~\AA, AIA 171~\AA\ with the point-specific epipolar loci, and the co-aligned \Ha\ centroid-velocity proxy map.
All retained stereoscopic tie points for each epoch are shown, with color indicating radial height above the adopted photosphere on the same 0.58--4.17~Mm scale as Figure~\ref{fig:stereo}.
Light-gray canvas labeled ``outside H-alpha coverage'' identifies projected locations beyond the available \Ha\ field rather than zero-valued Doppler data.
The panels extend the sampling into the later Stage~1, inter-stage, and Stage~2 morphology where available and visualize low-lying, inclined, and locally arched bright strands.
The measurements are independent epoch-by-epoch tie points rather than one continuously tracked feature.}
{fig:supp_stereo_b}

\subsection{Multi-layer spectral evolution at AIA cadence}
\label{app:multilayer}

Figure~\ref{fig:supp_multilayer} shows the multi-layer spatial evolution, including the precursor, the two releases, the k2 expansion during the Stage~2 decay, and the late spectral evolution.
The figure shows 15 selected IRIS rasters (R00, R02, R04, R07, R10, R13, R15, R19, R20, R22, R25, R28, R31, R39, and R44), each paired with the AIA 171~\AA\ exposure nearest the raster midpoint.

\paperanimation{MovieS2_FigureS4_multilayer.mp4}{\figsmultilayerfile}{0.91}
{Spatial evolution of coronal, TR, and chromospheric diagnostics in 15 selected IRIS rasters. From top to bottom: AIA 171~\AA; \MgII\ k3 and k2 intensities; k3 and k2 Doppler proxies; the observed \MgII\ k width proxy; and \SiIV\ 1394~\AA\ integrated intensity. Each column is labeled with its zero-based raster number and midpoint time; AIA uses the exposure nearest that midpoint. Every row retains the same cropped raster-scanned field, and an IRIS column is assembled during a scan rather than acquired instantaneously. Each diagnostic has a fixed scale across all columns. The k3 and k2 intensity rows share one color bar, and their Doppler rows share another; positive Doppler values denote redshift. AIA uses a square-root intensity display, and the \SiIV\ maps are in corrected DN~\AA. The width proxy is the FWHM-equivalent width derived from the observed profile's second moment, not a directly measured nonthermal velocity. Cyan outlines mark the AIA event region, teal outlines the IRIS-covered upper ROI, and yellow-green contours mark k2 intensity above 40~DN within that ROI on the k2-intensity, k2-Doppler, and width rows. The narrow amber and teal bars above the AIA row identify Stages~1 and~2. The compact k2 precursor is visible in row~(c), R02; its observed profiles are shown in Figure~\ref{fig:supp_profiles}.}
{fig:supp_multilayer}

\subsection{Observed \MgII\ profiles at the precursor site}
\label{app:profiles}

Figure~\ref{fig:supp_profiles} shows the observed k profiles at the precursor site. For the thick illustrative curves, profile shape is compared jointly in the two lines after amplitude normalization used only for selecting the central subset. The subset is the most compact neighborhood containing $\min(N,\max(3,\lceil N/2\rceil))$ of the $N$ usable paired profiles. The mean is then calculated from their original, unnormalized intensities. All usable individual profiles remain visible as thin curves. This display selection does not affect the regional central-reversal statistics.

\paperfigure{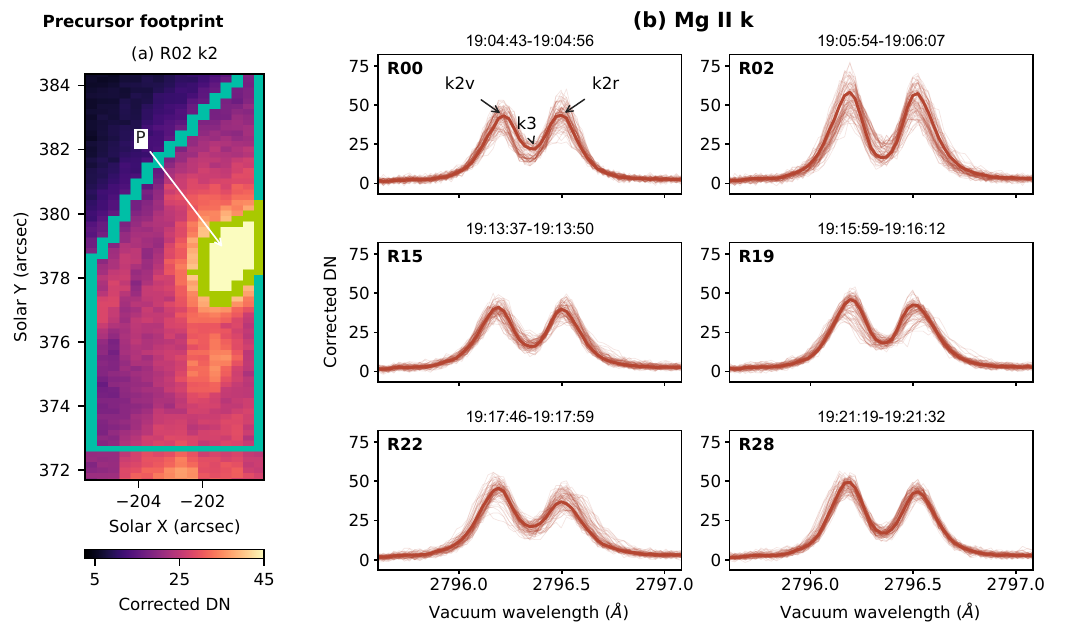}{0.99}
{Observed \MgII\ profiles at the single compact precursor site. (a) The R02 k2 map; the outlined source footprint is the connected bright feature inside the existing k2 contour within the upper ROI. This footprint is held fixed in the registered coordinate system for the spectral examples. (b) \MgII\ k profiles for R00, R02, R15, R19, R22, and R28, arranged in chronological order from left to right and top to bottom. Thin curves show every usable observed profile within the footprint at that raster. Thick curves are arithmetic means of the profiles in a fixed-rule central part of the shape distribution, evaluated in the original intensity units; 44 of the 87 profiles contribute to each thick curve. The subset selection still uses the joint h and k profile shapes. Labels give the exposure-time ranges of the sampled pixels. This illustrative subset does not replace the per-profile ratio medians in Figure~\ref{fig:timing}(c), and the spectra do not follow individual plasma elements.}
{fig:supp_profiles}

\subsection{Optical-flow diagnostics}
\label{app:of}

The TV-L1 optical-flow correlation gain is $r_{\rm after}-r_{\rm before}$, where $r$ is the Pearson correlation coefficient between consecutive \Ha\ line-center images evaluated over the same region before and after remapping. Both the residual and the gain respond to coherent apparent motion as well as to feature appearance or disappearance, intensity evolution, residual seeing or registration errors, and mismatch of the optical-flow model. The first two image pairs are excluded because residual FOV jumps dominate their inferred motion.

\subsection{Statistical tests for the cross-ridge red--blue separation}
\label{app:stats}

The permutation null hypothesis is that the red- and blueshift labels are unrelated to the signed cross-ridge offset $d_{\rm cr}$. The sampled offsets and the sizes of the redshifted, blueshifted, and unselected groups are kept fixed, the labels are randomly permuted, and $\Delta d_{\rm cr}$ is recomputed for 400 permutations. The two-sided Monte Carlo $p$-value is $[1+N(|\Delta d_{{\rm cr},{\rm perm}}|\geq|\Delta d_{{\rm cr},{\rm obs}}|)]/(400+1)$.
Stage-level confidence intervals are obtained from 20,000 bootstrap resamples of complete rasters, and leave-one-raster-out calculations test whether the stage contrast is controlled by a single raster.
The bootstrap resamples complete rasters and does not explicitly model temporal correlation between adjacent rasters.

At the adopted $3~\kms$ threshold, the stage medians of the redshifted and blueshifted positions are $-0.719\arcsec$ and $+0.751\arcsec$ for Stage~1 (R11--R17), and $-0.185\arcsec$ and $+0.704\arcsec$ for Stage~2 (R20--R25). The corresponding separations are $1.47\arcsec$ and $0.89\arcsec$, with 95\% raster-bootstrap intervals of $1.16$--$1.63\arcsec$ and $0.41$--$1.22\arcsec$, respectively. The Stage~1 minus Stage~2 difference is $0.58\arcsec$, with a 95\% interval of $0.05$--$1.13\arcsec$. These intervals reflect the adopted raster resampling and do not encompass all temporal or systematic uncertainties.

\subsection{Robustness checks for the \MgII\ diagnostics}
\label{app:robust}

We tested the cross-ridge separation against a uniform offset of the adopted k3 velocity zero point while keeping the spatial regions and ridge geometry fixed. For offsets from \(-5\) to \(+5~\mathrm{km\,s^{-1}}\), every stage-level comparison that can be evaluated retains a larger Stage~1 than Stage~2 separation. At the larger offsets, however, the red- or blueshifted samples become sparse and some rasters no longer contain enough samples for the permutation test, so the number of individually significant rasters depends on the zero point.

We also tested the \MgII\ h3/h2 and k3/k2 evolution with a fixed spatial footprint defined by the intersection of the adopted ridge regions. The all-raster and R11--R25 intersections are identical and contain 452 spatial pixels. The fixed-footprint curves preserve the Stage~1 maximum, elevated inter-stage ratios, the local Stage~2 rebound, and the subsequent decline. Their Spearman correlations with the time-dependent ridge-region curves are 0.994 for h3/h2 and 0.991 for k3/k2. This test controls the changing footprint, although the set of measurable profiles still varies from raster to raster.

\subsection{Pre-event comparison before Stage~2}
\label{app:pre_event_r19}

Table~\ref{tab:pre_event_r19} lists the pre-event statistics and the R19 measurements used in Section~\ref{sec:stage2_results}. The reference sample contains R00, R01, R03, R04, R05, and R06, with R02 excluded. The sample standard deviation uses $N-1$ in the denominator. R19 spans 19:15:39--19:16:12~UT and is the last complete raster before the Stage~2 onset at 19:16:31~UT. R20 spans 19:16:14--19:16:48~UT and crosses that onset, so it is not used as the fully pre-onset comparison.

\begin{deluxetable*}{llrrr}
\tablewidth{0pt}
\tabletypesize{\footnotesize}
\tablecaption{Pre-event statistics and the last complete IRIS raster before Stage~2.\label{tab:pre_event_r19}}
\tablehead{\colhead{Diagnostic} & \colhead{Sampling region} & \colhead{Pre-event mean} & \colhead{Sample SD} & \colhead{R19}}
\startdata
k3 redshift signal & IRIS-covered upper ROI & 214.325 & 64.429 & 210.730 \\
k3 blueshift signal & IRIS-covered upper ROI & 80.254 & 11.015 & 237.982 \\
h3/h2 & Ridge region & 0.4122 & 0.0242 & 0.5422 \\
h3/h2 & Broader upper ROI & 0.4985 & 0.0160 & 0.5052 \\
k3/k2 & Ridge region & 0.4212 & 0.0264 & 0.5584 \\
k3/k2 & Broader upper ROI & 0.5100 & 0.0189 & 0.5243 \\
\SiIV\ 1394~\AA & IRIS-covered upper ROI & $3.717\times10^4$ & $1.400\times10^4$ & $3.370\times10^5$ \\
\SiIV\ 1402~\AA & IRIS-covered upper ROI & $3.796\times10^4$ & $6.141\times10^3$ & $1.528\times10^5$ \\
\enddata
\tablecomments{The six pre-event rasters are R00, R01, R03, R04, R05, and R06. These are unnormalized diagnostic values, not the R00-normalized values displayed in Figure~\ref{fig:timing}. The Doppler signals are sums of velocity excess, with units of $\kms$, not mean velocities or mass fluxes. The central-reversal ratios are dimensionless. The \SiIV\ values are regional sums of the positive continuum-subtracted line signal integrated over velocity, in $(\mathrm{DN\,s^{-1}})(\mathrm{km\,s^{-1}})$; they are distinct from the corrected DN~\AA\ map quantity in Figure~\ref{fig:supp_multilayer}. The ridge and broader upper-ROI samples overlap. Sample SD measures pre-event raster-to-raster variability, not a confidence interval.}
\end{deluxetable*}


\begin{thebibliography}{}
\expandafter\ifx\csname natexlab\endcsname\relax\def\natexlab#1{#1}\fi
\providecommand{\url}[1]{\href{#1}{#1}}
\providecommand{\dodoi}[1]{doi:~\href{http://doi.org/#1}{\nolinkurl{#1}}}
\providecommand{\doeprint}[1]{\href{http://ascl.net/#1}{\nolinkurl{http://ascl.net/#1}}}
\providecommand{\doarXiv}[1]{\href{https://arxiv.org/abs/#1}{\nolinkurl{https://arxiv.org/abs/#1}}}

\bibitem[{Berghmans {et~al.}(2021)Berghmans, Auchère, Long, Soubrié, Mierla, Zhukov, Schühle, Antolin, Harra, Parenti, Podladchikova, Cuadrado, Buchlin, Dolla, Verbeeck, Gissot, Teriaca, Haberreiter, Katsiyannis, Rodriguez, Kraaikamp, Smith, Stegen, Rochus, Halain, Jacques, Thompson, \& Inhester}]{berghmans_extreme-uv_2021}
Berghmans, D., Auchère, F., Long, D.~M., {et~al.} 2021, A\&A, 656, L4, \dodoi{10.1051/0004-6361/202140380}

\bibitem[{Boerner {et~al.}(2012)Boerner, Edwards, Lemen, Rausch, Schrijver, Shine, Shing, Stern, Tarbell, Title, Wolfson, Soufli, Spiller, Gullikson, McKenzie, Windt, Golub, Podgorski, Testa, \& Weber}]{Boerner2012}
Boerner, P., Edwards, C., Lemen, J., {et~al.} 2012, SoPh, 275, 41, \dodoi{10.1007/s11207-011-9804-8}

\bibitem[{Carlsson {et~al.}(2015)Carlsson, Leenaarts, \& De~Pontieu}]{carlsson_what_2015}
Carlsson, M., Leenaarts, J., \& De~Pontieu, B. 2015, ApJL, 809, L30, \dodoi{10.1088/2041-8205/809/2/L30}

\bibitem[{Chen {et~al.}(2021)Chen, Przybylski, Peter, Tian, Auchère, \& Berghmans}]{chen_transient_2021}
Chen, Y., Przybylski, D., Peter, H., {et~al.} 2021, A\&A, 656, L7, \dodoi{10.1051/0004-6361/202140638}

\bibitem[{Chitta {et~al.}(2023)Chitta, Zhukov, Berghmans, Peter, Parenti, Mandal, Aznar~Cuadrado, Schühle, Teriaca, Auchère, Barczynski, Buchlin, Harra, Kraaikamp, Long, Rodriguez, Schwanitz, Smith, Verbeeck, \& Seaton}]{chitta_picoflare_2023}
Chitta, L.~P., Zhukov, A.~N., Berghmans, D., {et~al.} 2023, Science, 381, 867, \dodoi{10.1126/science.ade5801}

\bibitem[{Chitta {et~al.}(2025)Chitta, Huang, D’Amicis, Calchetti, Zhukov, Kraaikamp, Verbeeck, Aznar~Cuadrado, Hirzberger, Berghmans, Horbury, Solanki, Owen, Harra, Peter, Schühle, Teriaca, Louarn, Livi, Giunta, Hassler, \& Wang}]{chitta_coronal_2025}
Chitta, L.~P., Huang, Z., D’Amicis, R., {et~al.} 2025, A\&A, 694, A71, \dodoi{10.1051/0004-6361/202452737}

\bibitem[{De Pontieu {et~al.}(2014)De Pontieu, Title, Lemen, Kushner, Akin, Allard, Berger, Boerner, Cheung, Chou, Drake, Duncan, Freeland, Heyman, Hoffman, Hurlburt, Lindgren, Mathur, Rehse, Sabolish, Seguin, Schrijver, Tarbell, Wülser, Wolfson, Yanari, Mudge, Nguyen-Phuc, Timmons, Van~Bezooijen, Weingrod, Brookner, Butcher, Dougherty, Eder, Knagenhjelm, Larsen, Mansir, Phan, Boyle, Cheimets, DeLuca, Golub, Gates, Hertz, McKillop, Park, Perry, Podgorski, Reeves, Saar, Testa, Tian, Weber, Dunn, Eccles, Jaeggli, Kankelborg, Mashburn, Pust, Springer, Carvalho, Kleint, Marmie, Mazmanian, Pereira, Sawyer, Strong, Worden, Carlsson, Hansteen, Leenaarts, Wiesmann, Aloise, Chu, Bush, Scherrer, Brekke, Martinez-Sykora, Lites, McIntosh, Uitenbroek, Okamoto, Gummin, Auker, Jerram, Pool, \& Waltham}]{depontieu_interface_2014}
De Pontieu, B., Title, A.~M., Lemen, J.~R., {et~al.} 2014, SoPh, 289, 2733, \dodoi{10.1007/s11207-014-0485-y}

\bibitem[{Freeland \& Handy(2012)}]{Freeland2012}
Freeland, S.~L., \& Handy, B.~N. 2012, {SolarSoft}: {Programming} and {Data} {Analysis} {Environment} for {Solar} {Physics}

\bibitem[{Hannah \& Kontar(2012)}]{Hannah2012}
Hannah, I.~G., \& Kontar, E.~P. 2012, A\&A, 539, A146, \dodoi{10.1051/0004-6361/201117576}

\bibitem[{Huang {et~al.}(2023)Huang, D'Anna, \& Wang}]{huang_statistical_2023}
Huang, N., D'Anna, S., \& Wang, H. 2023, ApJL, 946, L17, \dodoi{10.3847/2041-8213/acc0f1}

\bibitem[{Huang \& Wang(2025)}]{huang_high-resolution_2025}
Huang, N., \& Wang, H. 2025, ApJ, 990, 77, \dodoi{10.3847/1538-4357/adf226}

\bibitem[{Kumar {et~al.}(2015)Kumar, Yurchyshyn, Wang, \& Cho}]{kumar_formation_2015}
Kumar, P., Yurchyshyn, V., Wang, H., \& Cho, K.-S. 2015, ApJ, 809, 83, \dodoi{10.1088/0004-637X/809/1/83}

\bibitem[{Lee {et~al.}(2025)Lee, Longcope, Youn, Panesar, Huang, \& Wang}]{lee_fine_2025}
Lee, J., Longcope, D., Youn, J., {et~al.} 2025, ApJL, 992, L23, \dodoi{10.3847/2041-8213/ae0df2}

\bibitem[{Leenaarts {et~al.}(2015)Leenaarts, Carlsson, \& Rouppe van~der Voort}]{leenaarts_fibrils_2015}
Leenaarts, J., Carlsson, M., \& Rouppe van~der Voort, L. 2015, ApJ, 802, 136, \dodoi{10.1088/0004-637X/802/2/136}

\bibitem[{Leenaarts {et~al.}(2013)Leenaarts, Pereira, Carlsson, Uitenbroek, \& De~Pontieu}]{leenaarts_formation_2013}
Leenaarts, J., Pereira, T. M.~D., Carlsson, M., Uitenbroek, H., \& De~Pontieu, B. 2013, ApJ, 772, 90, \dodoi{10.1088/0004-637X/772/2/90}

\bibitem[{Lemen {et~al.}(2012)Lemen, Title, Akin, Boerner, Chou, Drake, Duncan, Edwards, Friedlaender, Heyman, Hurlburt, Katz, Kushner, Levay, Lindgren, Mathur, McFeaters, Mitchell, Rehse, Schrijver, Springer, Stern, Tarbell, Wuelser, Wolfson, Yanari, Bookbinder, Cheimets, Caldwell, Deluca, Gates, Golub, Park, Podgorski, Bush, Scherrer, Gummin, Smith, Auker, Jerram, Pool, Soufli, Windt, Beardsley, Clapp, Lang, \& Waltham}]{lemen_atmospheric_2012}
Lemen, J.~R., Title, A.~M., Akin, D.~J., {et~al.} 2012, SoPh, 275, 17, \dodoi{10.1007/s11207-011-9776-8}

\bibitem[{Li {et~al.}(2018)Li, Li, \& Ning}]{li_spectroscopic_2018}
Li, D., Li, L., \& Ning, Z. 2018, MNRAS, 479, 2382, \dodoi{10.1093/mnras/sty1712}

\bibitem[{Li {et~al.}(2016)Li, Zhang, Peter, Priest, Chen, Guo, Chen, \& Mackay}]{li_magnetic_2016}
Li, L., Zhang, J., Peter, H., {et~al.} 2016, Nature Physics, 12, 847, \dodoi{10.1038/nphys3768}

\bibitem[{Longcope \& Klaassen(2025)}]{longcope_alfven_2025}
Longcope, D., \& Klaassen, P. 2025, ApJ, 989, 152, \dodoi{10.3847/1538-4357/adf0ef}

\bibitem[{Mandal {et~al.}(2023)Mandal, Peter, Chitta, Solanki, Aznar~Cuadrado, Schühle, Teriaca, Martínez-Sykora, Berghmans, Auchère, Parenti, Zhukov, Buchlin, Verbeeck, Kraaikamp, Rodriguez, Long, Barczynski, Pelouze, \& Smith}]{mandal_evolution_2023}
Mandal, S., Peter, H., Chitta, L.~P., {et~al.} 2023, A\&A, 678, L5, \dodoi{10.1051/0004-6361/202347343}

\bibitem[{Müller {et~al.}(2020)Müller, St.~Cyr, Zouganelis, Gilbert, Marsden, Nieves-Chinchilla, Antonucci, Auchère, Berghmans, Horbury, Howard, Krucker, Maksimovic, Owen, Rochus, Rodriguez-Pacheco, Romoli, Solanki, Bruno, Carlsson, Fludra, Harra, Hassler, Livi, Louarn, Peter, Schühle, Teriaca, del Toro~Iniesta, Wimmer-Schweingruber, Marsch, Velli, De~Groof, Walsh, \& Williams}]{muller_solar_2020}
Müller, D., St.~Cyr, O.~C., Zouganelis, I., {et~al.} 2020, A\&A, 642, A1, \dodoi{10.1051/0004-6361/202038467}

\bibitem[{Panesar {et~al.}(2018)Panesar, Sterling, Moore, Tiwari, Pontieu, \& Norton}]{panesar_iris_2018}
Panesar, N.~K., Sterling, A.~C., Moore, R.~L., {et~al.} 2018, ApJL, 868, L27, \dodoi{10.3847/2041-8213/aaef37}

\bibitem[{Panesar {et~al.}(2025)Panesar, Sterling, Moore, Tiwari, Berghmans, Zhukov, Mierla, Verbeeck, \& Stegen}]{panesar_buildup_2025}
---. 2025, ApJ, 994, 164, \dodoi{10.3847/1538-4357/ae0d90}

\bibitem[{Pariat {et~al.}(2009)Pariat, Antiochos, \& DeVore}]{Pariat2009}
Pariat, E., Antiochos, S.~K., \& DeVore, C.~R. 2009, ApJ, 691, 61, \dodoi{10.1088/0004-637X/691/1/61}

\bibitem[{Pariat {et~al.}(2015)Pariat, Dalmasse, DeVore, Antiochos, \& Karpen}]{Pariat2015}
Pariat, E., Dalmasse, K., DeVore, C.~R., Antiochos, S.~K., \& Karpen, J.~T. 2015, A\&A, 573, A130, \dodoi{10.1051/0004-6361/201424209}

\bibitem[{Pariat {et~al.}(2016)Pariat, Dalmasse, DeVore, Antiochos, \& Karpen}]{Pariat2016}
---. 2016, A\&A, 596, A36, \dodoi{10.1051/0004-6361/201629109}

\bibitem[{Pereira {et~al.}(2016)Pereira, Rouppe van~der Voort, \& Carlsson}]{pereira_appearance_2016}
Pereira, T. M.~D., Rouppe van~der Voort, L., \& Carlsson, M. 2016, ApJ, 824, 65, \dodoi{10.3847/0004-637X/824/2/65}

\bibitem[{Petrova {et~al.}(2024)Petrova, Van~Doorsselaere, Berghmans, Parenti, Valori, \& Plowman}]{petrova_observations_2024}
Petrova, E., Van~Doorsselaere, T., Berghmans, D., {et~al.} 2024, A\&A, 687, A13, \dodoi{10.1051/0004-6361/202348799}

\bibitem[{Raouafi {et~al.}(2016)Raouafi, Patsourakos, Pariat, Young, Sterling, Savcheva, Shimojo, Moreno-Insertis, DeVore, Archontis, Török, Mason, Curdt, Meyer, Dalmasse, \& Matsui}]{raouafi_solar_2016}
Raouafi, N.~E., Patsourakos, S., Pariat, E., {et~al.} 2016, SSRv, 201, 1, \dodoi{10.1007/s11214-016-0260-5}

\bibitem[{Rees \& Semel(1979)}]{rees_line_1979}
Rees, D.~E., \& Semel, M.~D. 1979, A\&A, 74, 1

\bibitem[{Rochus {et~al.}(2020)Rochus, Auchère, Berghmans, Harra, Schmutz, Schühle, Addison, Appourchaux, Aznar~Cuadrado, Baker, Barbay, Bates, BenMoussa, Bergmann, Beurthe, Borgo, Bonte, Bouzit, Bradley, Büchel, Buchlin, Büchner, Cabé, Cadiergues, Chaigneau, Chares, Choque~Cortez, Coker, Condamin, Coumar, Curdt, Cutler, Davies, Davison, Defise, Del~Zanna, Delmotte, Delouille, Dolla, Dumesnil, Dürig, Enge, François, Fourmond, Gillis, Giordanengo, Gissot, Green, Guerreiro, Guilbaud, Gyo, Haberreiter, Hafiz, Hailey, Halain, Hansotte, Hecquet, Heerlein, Hellin, Hemsley, Hermans, Hervier, Hochedez, Houbrechts, Ihsan, Jacques, Jérôme, Jones, Kahle, Kennedy, Klaproth, Kolleck, Koller, Kotsialos, Kraaikamp, Langer, Lawrenson, Le~Clech, Lenaerts, Liebecq, Linder, Long, Mampaey, Markiewicz-Innes, Marquet, Marsch, Matthews, Mazy, Mazzoli, Meining, Meltchakov, Mercier, Meyer, Monecke, Monfort, Morinaud, Moron, Mountney, Müller, Nicula, Parenti, Peter, Pfiffner, Philippon, Phillips, Plesseria, Pylyser, Rabecki,
  Ravet-Krill, Rebellato, Renotte, Rodriguez, Roose, Rosin, Rossi, Roth, Rouesnel, Roulliay, Rousseau, Ruane, Scanlan, Schlatter, Seaton, Silliman, Smit, Smith, Solanki, Spescha, Spencer, Stegen, Stockman, Szwec, Tamiatto, Tandy, Teriaca, Theobald, Tychon, van Driel-Gesztelyi, Verbeeck, Vial, Werner, West, Westwood, Wiegelmann, Willis, Winter, Zerr, Zhang, \& Zhukov}]{rochus_solar_2020}
Rochus, P., Auchère, F., Berghmans, D., {et~al.} 2020, A\&A, 642, A8, \dodoi{10.1051/0004-6361/201936663}

\bibitem[{Rouppe van~der Voort {et~al.}(2015)Rouppe van~der Voort, De~Pontieu, Pereira, Carlsson, \& Hansteen}]{rouppe_van_der_voort_heating_2015}
Rouppe van~der Voort, L., De~Pontieu, B., Pereira, T. M.~D., Carlsson, M., \& Hansteen, V. 2015, ApJL, 799, L3, \dodoi{10.1088/2041-8205/799/1/L3}

\bibitem[{Rubio~da Costa \& Kleint(2017)}]{rubio_da_costa_parameter_2017}
Rubio~da Costa, F., \& Kleint, L. 2017, ApJ, 842, 82, \dodoi{10.3847/1538-4357/aa6eaf}

\bibitem[{Sainz~Dalda {et~al.}(2019)Sainz~Dalda, de~la Cruz~Rodríguez, De~Pontieu, \& Gošić}]{sainz_dalda_recovering_2019}
Sainz~Dalda, A., de~la Cruz~Rodríguez, J., De~Pontieu, B., \& Gošić, M. 2019, ApJL, 875, L18, \dodoi{10.3847/2041-8213/ab15d9}

\bibitem[{Sterling {et~al.}(2015)Sterling, Moore, Falconer, \& Adams}]{sterling_small-scale_2015}
Sterling, A.~C., Moore, R.~L., Falconer, D.~A., \& Adams, M. 2015, Nature, 523, 437, \dodoi{10.1038/nature14556}

\bibitem[{Tang {et~al.}(2025)Tang, Shen, Liu, Yao, Zhou, \& Li}]{tang_photospheric_2025}
Tang, Z., Shen, Y., Liu, D., {et~al.} 2025, ApJL, 990, L16, \dodoi{10.3847/2041-8213/adf0fe}

\bibitem[{Uitenbroek(2003)}]{uitenbroek_accuracy_2003}
Uitenbroek, H. 2003, ApJ, 592, 1225, \dodoi{10.1086/375736}

\bibitem[{Xue {et~al.}(2016)Xue, Yan, Cheng, Yang, Su, Kliem, Zhang, Liu, Bi, Xiang, Yang, \& Zhao}]{xue_observing_2016}
Xue, Z., Yan, X., Cheng, X., {et~al.} 2016, Nature Communications, 7, 11837, \dodoi{10.1038/ncomms11837}

\bibitem[{Yang {et~al.}(2023)Yang, Yan, Xue, Chen, Wang, Xu, \& Li}]{yang_observational_2023}
Yang, L., Yan, X., Xue, Z., {et~al.} 2023, ApJ, 945, 96, \dodoi{10.3847/1538-4357/acb6f6}

\bibitem[{Youn {et~al.}(2025)Youn, Lee, Jeong, Lee, Park, \& Moon}]{youn_can_2025}
Youn, J., Lee, H., Jeong, H.-J., {et~al.} 2025, A\&A, 695, A125, \dodoi{10.1051/0004-6361/202452304}

\bibitem[{Zhukov {et~al.}(2021)Zhukov, Mierla, Auchère, Gissot, Rodriguez, Soubrié, Thompson, Inhester, Nicula, Antolin, Parenti, Buchlin, Barczynski, Verbeeck, Kraaikamp, Smith, Stegen, Dolla, Harra, Long, Schühle, Podladchikova, Aznar~Cuadrado, Teriaca, Haberreiter, Katsiyannis, Rochus, Halain, Jacques, \& Berghmans}]{zhukov_stereoscopy_2021}
Zhukov, A.~N., Mierla, M., Auchère, F., {et~al.} 2021, A\&A, 656, A35, \dodoi{10.1051/0004-6361/202141010}

\end{thebibliography}
\end{document}